\documentclass[lettersize,journal]{IEEEtran}
\usepackage{amsmath,amsfonts}
\usepackage{algorithmic}
\usepackage{algorithm}
\usepackage{array}
\usepackage[caption=false,font=normalsize,labelfont=sf,textfont=sf]{subfig}
\usepackage{textcomp}
\usepackage{stfloats}
\usepackage{url}
\usepackage{verbatim}
\usepackage{graphicx}
\usepackage{cite}
\usepackage{booktabs}
\usepackage[most]{tcolorbox}
\usepackage{xcolor}
\usepackage{varwidth}
\usepackage{enumitem}
\usepackage{multirow}
\usepackage{wrapfig}
\usepackage{caption}
\definecolor{inputcolor}{RGB}{35,95,170}
\definecolor{outputcolor}{RGB}{20,130,80}
\definecolor{democolor}{RGB}{180,110,20}
\definecolor{notecolor}{RGB}{70,70,70}
\definecolor{varcolor}{RGB}{150,40,140}

\newcommand{\InputTag}{\textbf{\textcolor{inputcolor}{[Input]}}}
\newcommand{\OutputTag}{\textbf{\textcolor{outputcolor}{[Output]}}}
\newcommand{\DemoTag}{\textbf{\textcolor{democolor}{[Reference Trajectory]}}}

\newcommand{\Var}[1]{\textcolor{varcolor}{#1}}

\begin{document}

\title{Self-Distilled Reinforcement Learning for Co-Evolving Agentic Recommender Systems}


\author{Zongwei Wang,
        Min Gao*,
        Hongzhi Yin*,
        Junliang Yu,
        Tong Chen,\\
        Quoc Viet Hung Nguyen,
        Shazia Sadiq,
        and Tianrui Li%
\thanks{*Corresponding authors: Min Gao and Hongzhi Yin.}
\thanks{Zongwei Wang and Min Gao are with the School of Big Data and Software Engineering, Chongqing University, Chongqing, China (e-mail: zongwei@cqu.edu.cn; gaomin@cqu.edu.cn).}
\thanks{Junliang Yu and Quoc Viet Hung Nguyen are with the School of Information and Communication Technology, Griffith University, Brisbane, Australia (e-mail: junliang.yu@griffith.edu.au;henry.nguyen@griffith.edu.au).}
\thanks{Hongzhi Yin, Tong Chen, and Shazia Sadiq are with the School of Information Technology and Electrical Engineering, The University of Queensland, Brisbane, Australia (e-mail: h.yin1@uq.edu.au; tong.chen@uq.edu.au; shazia@eecs.uq.edu.au).}%
\thanks{Tianrui Li is with the School of Computing and Artificial Intelligence, Southwest Jiaotong University, Chengdu, China (e-mail: trli@swjtu.edu.cn).}
}




\markboth{Journal of \LaTeX\ Class Files,~Vol.~14, No.~8, August~2021}%
{Shell \MakeLowercase{\textit{et al.}}: A Sample Article Using IEEEtran.cls for IEEE Journals}


\maketitle

\begin{abstract}
Large language model-empowered agentic recommender systems (ARS) reformulate recommendation as a multi-turn interaction between a recommender agent and a user agent, enabling iterative preference elicitation and refinement beyond conventional one-shot prediction. However, existing ARS are mainly optimized in a Reflexion-style paradigm, where past interaction trajectories are stored as textual memory and retrieved as prompt context for later reasoning. Although this design allows agents to recall prior feedback and observations, the accumulated experience remains external to model parameters, leaving agents reliant on generic reasoning rather than progressively acquiring recommendation-specific decision-making ability through learning. Reinforcement learning (RL) therefore provides a natural way to internalize such interaction experience into parameter updates. Yet existing RL methods for ARS still mainly rely on predefined or judgment-based rewards, which suffer from two distinct limitations. First, their supervision is externally imposed or one-directional, rather than arising endogenously from the interaction between the recommender agent and the user agent. As a result, they cannot fully reflect the interactive nature that defines ARS. Second, they reduce a rich multi-turn interaction process to final outcomes, overlooking the dense supervision embedded throughout the trajectory. To this end, we propose CoARS, a self-distilled reinforcement learning framework for co-evolving agentic recommender systems. CoARS introduces two complementary learning schemes: interaction reward, which derives coupled task-level supervision for the recommender agent and the user agent from the same interaction trajectory, and self-distilled credit assignment, which converts historical trajectories into token-level credit signals under teacher-student conditioning. Experiments on multiple benchmarks show that CoARS consistently outperforms representative ARS baselines in both recommendation performance and user alignment.
\end{abstract}


\section{Introduction}
\IEEEPARstart{L}{arge} language models (LLMs) are opening up new possibilities for recommender systems beyond conventional one-shot prediction. In traditional recommendation pipelines, a model usually predicts a user’s preference from historical behaviors in a single pass, treating the user mainly as a source of static feedback signals such as clicks, ratings, or purchases~\cite{01kang2018self,10he2020lightgcn,21wang2023efficient}. By contrast, emerging LLM-powered agentic recommender systems (ARS) introduce a new paradigm that treats recommendation as an interactive decision-making process: a recommendation agent can generate proposals by reasoning over user history, while a user agent can provide explicit feedback (e.g., like or dislike) to guide subsequent revisions until an item is finally accepted~\cite{04cai2025agentic,02xu2025iagent,11zhang2024generative,26huang2025recommender}. Such a multi-turn process enables the system to anticipate user reactions, iteratively refine recommendations, and better align recommendation outcomes with user preferences.

\begin{figure*}[t]
	\centering	\includegraphics[width=0.95\linewidth]{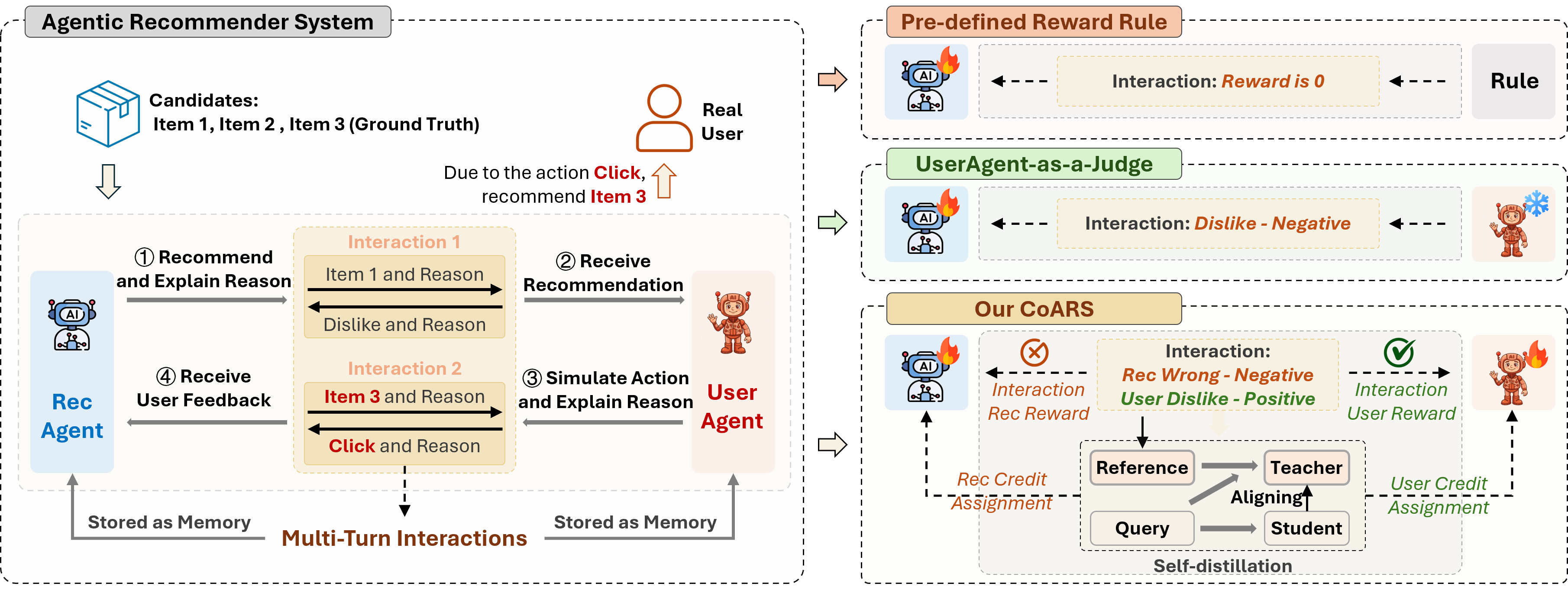}
    \caption{Comparison of evolution paradigms in agentic recommender systems. The left part shows the standard ARS pipeline with the Reflexion paradigm. The right part compares three RL paradigms.}
	\label{introduction}
\end{figure*}

Despite these advantages, existing ARS are still mainly optimized in a Reflexion-style paradigm~\cite{06shinn2023reflexion}. As illustrated in the left part of Fig.~\ref{introduction}, multi-turn interaction trajectories are stored as textual memory and retrieved as additional prompt context in later interactions~\cite{34tran2026entropy,02xu2025iagent,04cai2025agentic}. This design helps agents recall previous feedback and observations, but the accumulated experience remains outside the model parameters~\cite{38zhong2024memorybank,45xu2025mem}. Consequently, agents still rely largely on generic reasoning, rather than progressively developing reasoning tailored to recommendation generation and user decision making through parameter updates~\cite{39bao2023tallrec}. Due to this limitation, researchers have naturally turned to reinforcement learning (RL) for ARS to internalize interaction experience into model parameters~\cite{32yu2025dapo,33schulman2017proximal}. As illustrated in the right part of Fig.~\ref{introduction}, one line of work relies on predefined reward rules~\cite{08guo2025deepseek}, where interaction quality is assessed by hand-crafted criteria or external judgments~\cite{09nguyen2026amem4rec}. Another line adopts an agent-as-judge paradigm~\cite{07you2026agent}, in which one agent evaluates the output of another and provides judge-derived rewards for optimization~\cite{05liu2025recoworld}. These paradigms have provided practical foundations for applying RL to ARS.

However, we find that these reward paradigms still fall short of the interactive nature of ARS. Specifically, predefined rewards depend heavily on human priors and often generalize poorly across settings, while judge-based rewards typically optimize only the actor while keeping the judge fixed. Yet in ARS, the recommender agent and the user agent continuously influence and update each other within the interaction loop. A better-suited reward design should emerge from interaction feedback, so as to support the joint optimization of both agents, rather than being tied to manually specified verification criteria or improving only one agent against a fixed evaluator. Moreover, both paradigms collapse a rich multi-turn interaction process into outcome-level rewards, overlooking the dense supervision embedded in the trajectory itself. Beyond indicating whether a recommendation succeeds, the trajectory also reveals how both agents reason, respond, and adjust at intermediate steps. Based on these observations, we argue that reward design for ARS should be endogenous to interaction feedback and make full use of the dense supervision provided by interaction trajectories, so that the recommender agent and the user agent can co-evolve through the interaction loop.

In this paper, we propose self-distilled reinforcement learning for \textbf{Co}-evolving \textbf{A}gentic \textbf{R}ecommender \textbf{S}ystems, named CoARS, a framework built on two complementary learning schemes. For the former, we design an interaction reward that decomposes each interaction turn into coupled supervision for both agents, enabling each agent to receive its own learning signal and optimizing distinct reasoning behaviors within the same interaction. Unlike rewards derived from one-sided feedback alone, the proposed reward is jointly defined by the combined decision outcomes of the recommendation agent and the user agent, allowing the bidirectional optimization to capture not only whether the recommended item is correct, but also how the user responds to it. For the latter, we introduce a self-distilled credit assignment scheme that enables each agent to learn from its own history. Specifically, we rewrite diagnosed historical interaction trajectories into reference trajectories, which identify better and worse decisions and highlight more desirable reasoning paths in hindsight. We then cast the same agent into teacher and student modes under different contexts: the teacher is conditioned on both the original interaction context and the reference trajectories, while the student only sees the original interaction context. By assigning token-level credit based on teacher-student comparisons, historical reference trajectories are transformed from passive records into learning signals, allowing the agent to absorb past experience into its own reasoning process. In summary, our work makes the following contributions:

\begin{itemize}[leftmargin=12pt]
    \item We present a new perspective for ARS by formulating agent training as a co-evolutionary learning process, in which both the recommendation agent and the user agent are optimized from the shared interaction trajectories.

    \item We propose the self-distilled reinforcement learning framework for ARS, which consists of an interaction reward that converts interaction outcomes into bidirectional supervision for both agents, and a self-distilled credit assignment scheme that transforms historical trajectories into token-level signals for policy optimization.

    \item We conduct extensive experiments on multiple recommendation benchmarks and show that the proposed framework consistently improves recommendation quality and user-alignment quality.
\end{itemize}

The rest of this paper is structured as follows. Section \ref{Preliminary} provides the background on the relevant preliminaries. Section \ref{Module} presents the module design of our proposed CoARS. The experimental results and analysis are presented in Section \ref{Experiments}, and Section \ref{casestudy} provides a case study including the detailed prompt and output of CoARS. Section \ref{related} reviews the related work for the agentic recommender system. The paper is concluded in Section \ref{conclusion}.

\section{Preliminary}
\label{Preliminary}
\subsection{Agentic Recommender Systems}

Agentic recommender systems (ARS) span multiple architectural forms, including single-agent recommendation frameworks~\cite{26huang2025recommender,41wang2025ruleagent}, generic multi-agent orchestration pipelines\cite{40wang2024macrec}, and interactive dual-agent systems~\cite{04cai2025agentic}. This work studies the latter, where a recommender-side agent (RecAgent) and a user-side agent (UserAgent) engage in multi-turn interactions to support recommendation refinement. In this paradigm, RecAgent produces both a candidate decision and an explicit rationale, while the UserAgent evaluates the recommendation, returns behavioral feedback, and also provides explicit justification. Such a dual-agent architecture turns recommendation into an iterative feedback loop in which both sides can continuously condition the evolving interaction history.

Formally, let $\mathcal{U}$ denote the user set, and let $\mathcal{I}$ denote the item set. For user $u \in \mathcal{U}$, let $h_u$ denote the historical interaction sequence. Suppose that completing one recommendation for user $u$ requires at most $T_u$ rounds of interaction between the RecAgent and the UserAgent.

\paragraph{RecAgent process}
At turn $t$, the RecAgent takes as input the user history $h_u$, the candidate set $\mathcal{C}_t \subseteq \mathcal{I}$, and the current memory $\mathcal{M}_{t-1}$, and generates a recommendation message:
\begin{equation}
\mathbf{m}^{\mathrm{rec}}_t = (i_t, r_t) \sim \pi_{\mathrm{rec}}(\cdot \mid h_u, \mathcal{C}_t, \mathcal{M}_{t-1}),
\end{equation}
where $\pi_{\mathrm{rec}}$ denotes the RecAgent policy, $i_t \in \mathcal{C}_t$ is the recommended item, and $r_t$ is the corresponding recommendation rationale.

\paragraph{UserAgent process}
Conditioned on the recommendation message $\mathbf{m}^{\mathrm{rec}}_t$, the UserAgent returns a feedback message:
\begin{equation}
\mathbf{m}^{\mathrm{user}}_t = (a_t, s_t, j_t) \sim \pi_{\mathrm{user}}(\cdot \mid h_u, \mathbf{m}^{\mathrm{rec}}_t, \mathcal{M}_{t-1}, p_t),
\end{equation}
where $\pi_{\mathrm{user}}$ denotes the UserAgent policy, $a_t \in \{\texttt{click}, \texttt{star}, \texttt{skip}, \texttt{dislike}\}$ denotes the user action, $s_t \in [0,1]$ is the acceptance score reflecting the intensity of user response, and $j_t$ is the feedback rationale. Here, $p_t$ denotes auxiliary evidence, i.e., recommendation opinions provided by other users who have also interacted with the current item.

We associate each action with a score interval: $\texttt{click}$ corresponds to $s_t \in (0.8,1.0]$, $\texttt{star}$ to $s_t \in (0.5,0.8]$, $\texttt{skip}$ to $s_t \in (0.3,0.5]$, and $\texttt{dislike}$ to $s_t \in [0,0.3]$. These intervals characterize progressively weaker levels of user acceptance, ranging from immediate acceptance to explicit rejection.

\paragraph{Multi-turn interaction}
We denote one interaction turn as:
\begin{equation}
\tau_t = (h_u, \mathcal{C}_t, \mathcal{M}_{t-1}, \mathbf{m}^{\mathrm{rec}}_t, \mathbf{m}^{\mathrm{user}}_t).
\end{equation}

Accordingly, the full interaction process for user $u$ is represented as an ordered trajectory
\begin{equation}
\mathcal{E}_u = (\tau_t)_{t=1}^{T_u}.
\end{equation}
The interaction proceeds iteratively until the UserAgent outputs $\texttt{click}$. If no $\texttt{click}$ is produced within the maximum number of turns, the system returns the item associated with the highest acceptance score observed across all turns.

\subsection{The Evolution of Agentic Recommender Systems}
Current ARS research mainly relies on two evolution mechanisms: memory update and reinforcement learning. Although both improve the agent through accumulated interactions, they operate at different levels and exhibit distinct limitations.

\paragraph{Memory update}
The standard evolution mechanism in existing ARS is primarily realized through memory update based on the Reflexion-style paradigm~\cite{06shinn2023reflexion}. Such a design enables both RecAgent and UserAgent to condition future decisions on richer historical context, thereby improving consistency across turns and helping the agents recall prior successes, failures, and preference cues. Formally, after each interaction turn, the memory is updated by appending the current interaction record:
\begin{equation}
\mathcal{M}_t = \mathcal{M}_{t-1} \cup \{(\mathbf{m}^{\mathrm{rec}}_t, \mathbf{m}^{\mathrm{user}}_t)\}.
\end{equation}

In practice, the memory is serialized into natural-language records and injected into the prompts of both agents in the next turn. However, this form of evolution remains confined to the prompt level: historical interactions are reused as textual context rather than converted into direct parameter-level supervision. As a result, the benefit of interaction history depends on in-context conditioning, while the agent itself does not explicitly absorb such experience into its model parameters.

\paragraph{Reinforcement learning}
Beyond memory update, a natural way to evolve ARS is to optimize the agent with reinforcement learning (RL)~\cite{35song2026expanding,36hou2025treerl,46lee2023supervised,47song2025reward}. In general, RL aims to maximize the expected reward, which in ARS is typically derived either from human-defined reward functions or from an agent-as-judge paradigm that evaluates the quality of generated recommendations and provides supervision for optimization.

Formally, let $R_t$ denote the reward assigned to the agent output at interaction turn $\tau_t$. Let $x_t$ denote the input context of the agent at turn $t$, and let $\mathbf{m}_t$ denote the corresponding output generated by the policy. A standard reinforcement learning objective is to maximize the expected cumulative reward over users and their interaction trajectories:
\begin{equation}
J_{\mathrm{RL}}
=
\mathbb{E}_{(u,\mathcal{E}_u)\sim\mathcal{D}}
\left[
\sum_{t=1}^{T_u} R_t \log \pi(\mathbf{m}_t \mid x_t)
\right],
\end{equation}
where $\mathcal{D}$ denotes the training distribution over users and their interaction trajectories. Under this objective, outputs that receive higher rewards are assigned higher learning preference during optimization, and are therefore more likely to be reinforced by the policy. 

The existing RL paradigm for ARS evolution still suffers from some major limitations. In particular, it largely follows reinforcement learning with verifiable rewards (RLVR)~\cite{32yu2025dapo,37wen2025reinforcement}, where historical collaborative feedback is compressed into sparse scalar supervision. Such signals can indicate whether an interaction is globally good or bad, but provide little guidance on which specific reasoning path or feedback pattern is responsible for success or failure. Moreover, this reward design is typically based on manually designed or externally constructed criteria~\cite{08guo2025deepseek}, which makes it difficult to fully capture the endogenous and interaction-dependent nature of agent collaboration. It is also commonly coupled with one-sided optimization~\cite{07you2026agent}, where only one agent, e.g., the RecAgent, is treated as the learnable policy, while the other remains fixed. This is problematic because a fixed agent may continually produce biased or suboptimal feedback, thereby limiting the quality of collaborative learning.

In the next section, we introduce CoARS, which leverages historical collaborative feedback, without requiring additional external supervision, to derive both interaction rewards and self-distilled credit assignment, thereby enabling the co-evolution of RecAgent and UserAgent.

\section{Our Method: CoARS}
\label{Module}
\begin{figure*}[t]
	\centering	\includegraphics[width=0.95\linewidth]{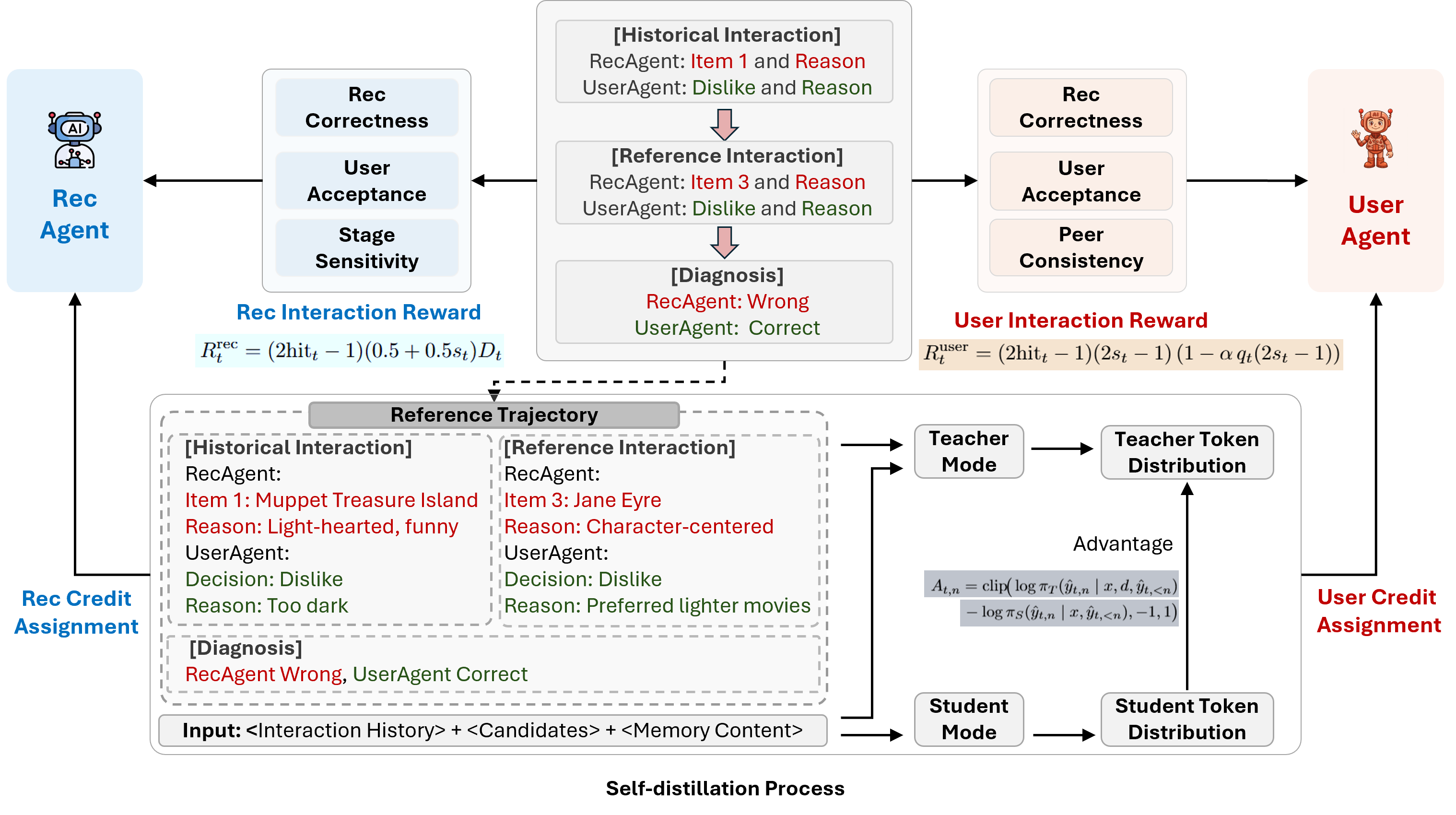}
    \caption{The framework of our CoARS.}
	\label{framework}
\end{figure*}

In ARS, RecAgent and UserAgent do not solve the same problem. RecAgent needs to reason about what to recommend and how to refine recommendations under feedback, while UserAgent needs to reason about how to interpret, evaluate, and respond to recommendations according to its preference state. Our key insight is that historical interactions in ARS contain much richer supervision than these existing mechanisms exploit. Each interaction naturally produces coupled outputs and reasoning trajectories from both RecAgent and UserAgent, thereby revealing whether the two agents are collaborating effectively. This makes it possible to naturally transform a single interaction into richer mutual learning signals for both sides. As shown in Fig.~\ref{framework}, we propose two key components in CoARS: the interaction reward that provides collaborative outcome-level supervision, and a self-distilled credit assignment scheme that leverages diagnosed historical reasoning trajectories to generate token-level supervision for both agents. Next, we present these two components in detail.


\subsection{Interaction Reward}

A successful interaction in ARS depends on the joint behavior of both agents: the RecAgent should produce a correct and persuasive recommendation, and the UserAgent should provide a response that is consistent with its actual preference state. Therefore, one interaction should naturally yield two coupled reward signals, one for each side.

\paragraph{RecAgent interaction reward}
For the RecAgent, the interaction reward is designed to jointly capture three aspects: recommendation correctness, user acceptance strength, and interaction-stage sensitivity. Specifically, the reward should distinguish correct from incorrect recommendations, reflect how strongly the UserAgent accepts the recommended item, and impose larger penalties on mistakes made at later interaction turns. Based on these considerations, we define the RecAgent interaction reward at turn $t$ as
\begin{equation}
R_t^{\mathrm{rec}} = (2\mathrm{hit}_t - 1)(0.5 + 0.5s_t)D_t.
\end{equation}
Here, $\mathrm{hit}_t \in \{0,1\}$ is the correctness indicator, where $\mathrm{hit}_t=1$ means that the recommended item matches the ground-truth item and $\mathrm{hit}_t=0$ otherwise. Therefore, the term $(2\mathrm{hit}_t-1)$ determines the reward direction, making the reward positive for correct recommendations and negative for incorrect ones. The variable $s_t \in [0,1]$ denotes the acceptance strength returned by the UserAgent. The term $(0.5+0.5s_t)$ then scales the reward magnitude according to how strongly the recommendation is accepted, while keeping the magnitude in a stable range. The factor $D_t$ captures the stage of the interaction. Before the correct item is found, $D_t$ increases exponentially with the interaction depth, so that mistakes made at later turns receive progressively larger penalties. After the correct item is identified, we fix $D_t = 1$ to prevent correct recommendations from being further magnified by stage-dependent scaling. As a result, successful recommendations are rewarded solely based on their correctness and acceptance strength.

\paragraph{UserAgent interaction reward}
For the UserAgent, the interaction reward is designed to capture three aspects simultaneously: recommendation correctness, user response direction and strength, and peer similarity. Specifically, the reward should encourage the UserAgent to accept correct recommendations and reject incorrect ones, while using peer similarity as an auxiliary signal to modulate the reward magnitude. Intuitively, a larger peer similarity indicates that peers who have interacted with the same item are more likely to share similar preferences with the current user. Therefore, making the same acceptance decision as highly similar peers is easier, whereas making a rejection decision against such similar peers is harder and thus more informative. We therefore define the UserAgent interaction reward at turn $t$ as:
\begin{equation}
R_t^{\mathrm{user}} =
(2\mathrm{hit}_t - 1)(2s_t - 1)\left(1 - \alpha\, q_t (2s_t - 1)\right),
\end{equation}
where $\mathrm{hit}_t \in \{0,1\}$ indicates whether the recommendation is correct, $s_t \in [0,1]$ denotes the acceptance strength returned by the UserAgent, and $q_t \in [-1,1]$ is the peer similarity score. The coefficient $\alpha$ controls the influence of peer similarity and is set to $0.1$ in practice to keep this effect moderate. In our method, $q_t$ is measured by the similarity between user embeddings learned from a conventional recommendation model such as SASRec~\cite{01kang2018self}. For better interpretability, Table~\ref{tab:user_reward_cases} summarizes the UserAgent reward under the eight possible cases defined by recommendation correctness, user response, and the sign of peer similarity.

\begin{table}[t]
\centering
\scriptsize
\setlength{\tabcolsep}{3pt}
\caption{Case-by-case behavior of the UserAgent interaction reward.}
\label{tab:user_reward_cases}
\begin{tabular}{c c c c c}
\toprule
$\mathrm{Hit}_t$ & User Response & Peer Type & Reward Sign & Effect of $q_t$ \\
\midrule
1 & Accept ($s_t>0.5$) & similar ($q_t>0$)    & $+$ & less positive \\
1 & Accept ($s_t>0.5$) & dissimilar ($q_t<0$) & $+$ & more positive \\
0 & Accept ($s_t>0.5$) & similar ($q_t>0$)    & $-$ & less negative \\
0 & Accept ($s_t>0.5$) & dissimilar ($q_t<0$) & $-$ & more negative \\
1 & Reject ($s_t<0.5$) & similar ($q_t>0$)    & $-$ & more negative \\
1 & Reject ($s_t<0.5$) & dissimilar ($q_t<0$) & $-$ & less negative \\
0 & Reject ($s_t<0.5$) & similar ($q_t>0$)    & $+$ & more positive \\
0 & Reject ($s_t<0.5$) & dissimilar ($q_t<0$) & $+$ & less positive \\
\bottomrule
\end{tabular}
\end{table}

It should be noted that in our implementation, if the recommendation is incorrect but the UserAgent still gives a positive reaction such as click or star, this interaction is excluded from RL updates. The reason is that such a case may simply indicate that the recommended item is unobserved in the dataset, rather than that the RecAgent has made a genuinely bad recommendation. Therefore, these samples are not treated as reliable supervision for optimization.

\subsection{Self-distilled Credit Assignment}
Although interaction reward provides collaborative supervision to enable genuine bidirectional co-evolution, it still remains an outcome-level signal. It can tell the agents whether an interaction is globally better or worse, but it offers limited guidance on which specific reasoning steps should be encouraged or corrected. 
Therefore, beyond outcome-level reward, ARS requires a finer-grained optimization mechanism that can identify which parts of historical reasoning should be reinforced or corrected. Our key motivation is that historical interactions in ARS naturally contain such information: once an interaction has occurred, it already records what the agents decided, how they responded, and why the interaction was relatively successful or unsuccessful. As a result, these trajectories can naturally serve as feedback-rich references for future learning.

Based on this observation, we further introduce a self-distillation credit assignment scheme to provide token-level credit signals. The self-distillation approach~\cite{18shenfeld2026self,19hubotter2026reinforcement,20zhao2026self} is an elegant solution fit for ARS: Instead of relying on an external teacher model or manually annotated supervision, we reuse the agent's own diagnosed historical trajectories as privileged reference trajectories. Concretely, the same agent is cast into two modes: the student only sees the original interaction context, while the teacher additionally observes a reference trajectory constructed from a completed interaction record. This reference is formed naturally by diagnosing the historical trajectory with the ground-truth item and a retrospective explanation of why the interaction succeeded or failed, and it contains the interaction context, recommendation, rationale, user response, feedback rationale, target item, and diagnostic explanation. In this way, the teacher represents how the same agent would reason when equipped with richer historical guidance, while the student is optimized to favor tokens more strongly supported by the teacher under the original interaction condition. Thus, already-occurred interactions are naturally transformed into token-level supervision for refinement.

Formally, let $d_t$ denote the diagnostic reference constructed from a historical interaction trajectory:
\begin{equation}
d_t = \bigl(\mathbf{m}^{\mathrm{rec}}_t,\mathbf{m}^{\mathrm{user}}_t,\mathbf{m}^{\mathrm{rec}*}_t,\mathbf{m}^{\mathrm{user}*}_t,e_t\bigr),
\end{equation}
where $\mathbf{m}^{\mathrm{rec}}_t$ and $\mathbf{m}^{\mathrm{user}}_t$ denote the original recommendation and user feedback messages, respectively. The corrected recommendation message $\mathbf{m}^{\mathrm{rec}*}_t$ is obtained by preserving the original interaction context, substituting the recommended item with the ground-truth item, and re-generating the recommendation rationale accordingly. Based on $\mathbf{m}^{\mathrm{rec}*}_t$, the corrected user feedback message $\mathbf{m}^{\mathrm{user}*}_t$ is then generated by prompting the user agent to produce the appropriate response together with its justification. Finally, $e_t$ is a diagnostic indicator specifying whether the original recommendation and user response are correct.

Let $x$ denote the original interaction context of the current agent. The student policy is defined as $\pi_S(\cdot \mid x)$, while the teacher policy additionally conditions on the reference trajectory, i.e., $\pi_T(\cdot \mid x,d)$. Given an on-policy student rollout at turn $t$, denoted by $\hat{y}_t=(\hat{y}_{t,1},\ldots,\hat{y}_{t,|\hat{y}_t|})$, we follow a concurrent work~\cite{43yang2026self} and compute a token-level diagnostic advantage by comparing the teacher and student log-probabilities on each sampled token:
\begin{equation}
\begin{aligned}
A_{t,n} = \mathrm{clip}\!\big(&
\log \pi_T(\hat{y}_{t,n} \mid x,d,\hat{y}_{t,<n}) \\
&- \log \pi_S(\hat{y}_{t,n} \mid x,\hat{y}_{t,<n}),
-1, 1
\big).
\end{aligned}
\end{equation}
Here, $A_{t,n}$ measures how much more strongly the teacher supports the $n$-th sampled token at turn $t$ than the student does. We further clip $A_{t,n}$ to avoid excessively large token-level advantages, which helps stabilize training while preserving the direction of the diagnostic signal.

To make the distinction between the two modes more concrete, we use a unified prompt template to illustrate the self-distillation mechanism. The \textbf{student mode} takes only the original interaction context as input and produces its response under the standard reasoning condition. The \textbf{teacher mode} uses the same prompt backbone, but additionally receives a diagnosed reference block constructed from historical interaction records. This block summarizes the original interaction, its diagnosed weaknesses, and a better reference reasoning path. In this way, the teacher represents how the same agent would reason when equipped with privileged retrospective guidance, while the student is trained to approximate this behavior under the original interaction condition.

\begin{tcolorbox}[
    title={Unified Prompt Template for Teacher Mode},
    colback=white,
    colframe=black,
    breakable,
    sharp corners
]
\ttfamily\small

<original interaction context>\\
<historical interaction memory>\\
<current reasoning task>

\vspace{0.5em}

\textcolor{democolor}{[Reference Trajectory]}\\
\textcolor{democolor}{Original interaction: <what was} 
\textcolor{democolor}{generated in the historical interaction>}\\
\textcolor{democolor}{Original reasoning: <how the agent reasoned in that interaction>}\\
\textcolor{democolor}{Observed feedback: <how the counterpart responded>}\\
\textcolor{democolor}{Interaction diagnosis: <why the interaction was relatively good or bad>}\\
\textcolor{democolor}{Reference reasoning: <a better reasoning path inferred retrospectively>}\\
\textcolor{democolor}{Reference response: <a better aligned decision or reaction>}\\
\textcolor{democolor}{Reference explanation: <why this revised reasoning/response is better>}\\
\\
\textcolor{democolor}{Instruction: use the reference trajectory as additional context.}\\
\textcolor{democolor}{Instruction: after the reference block, continue with the same task and output format as the original prompt.}

\vspace{0.5em}

<teacher-generated reasoning output>

\end{tcolorbox}

\subsection{Overall Training Objectives}

By combining interaction reward with self-distillation, CoARS optimizes RecAgent and UserAgent with separate but complementary objectives. In both cases, the interaction reward provides a shared task-level signal for all generated tokens, while the self-distillation advantage further assigns fine-grained token-level credit.

\paragraph{RecAgent objective}
For RecAgent, the overall objective is defined as
\begin{equation}
\begin{aligned}
J_{\mathrm{rec}}
=
\mathbb{E}_{(u,\mathcal{E}_u)\sim\mathcal{D}}
\Bigg[
\sum_{t=1}^{T_u}
\frac{1}{|y_t^{\mathrm{rec}}|}
\sum_{n=1}^{|y_t^{\mathrm{rec}}|}
\Big(
R_t^{\mathrm{rec}} + \lambda_{\mathrm{SD}}^{\mathrm{rec}} A_{t,n}^{\mathrm{rec}}
\Big) \\
\cdot
\log \pi_S^{\mathrm{rec}}(y_{t,n}^{\mathrm{rec}} \mid x_t^{\mathrm{rec}}, y_{t,<n}^{\mathrm{rec}})
\Bigg],
\end{aligned}
\end{equation}
where $y_t^{\mathrm{rec}}$ denotes the token sequence generated by RecAgent at turn $t$, $R_t^{\mathrm{rec}}$ is the interaction reward shared by all tokens in that turn, and $A_{t,n}^{\mathrm{rec}}$ is the token-level diagnostic advantage for the $n$-th token. The $\lambda_{\mathrm{SD}}^{\mathrm{rec}}$ controls the trade-off between these two terms.

\paragraph{UserAgent objective} For UserAgent, the overall objective is defined as
\begin{equation}
\begin{aligned}
J_{\mathrm{user}}
=
\mathbb{E}_{(u,\mathcal{E}_u)\sim\mathcal{D}}
\Bigg[
\sum_{t=1}^{T_u}
\frac{1}{|y_t^{\mathrm{user}}|}
\sum_{n=1}^{|y_t^{\mathrm{user}}|}
\Big(
R_t^{\mathrm{user}} + \lambda_{\mathrm{SD}}^{\mathrm{user}} A_{t,n}^{\mathrm{user}}
\Big) \\
\cdot
\log \pi_S^{\mathrm{user}}(y_{t,n}^{\mathrm{user}} \mid x_t^{\mathrm{user}}, y_{t,<n}^{\mathrm{user}})
\Bigg],
\end{aligned}
\end{equation}
where $y_t^{\mathrm{user}}$ denotes the token sequence generated by UserAgent at turn $t$, $R_t^{\mathrm{user}}$ is the interaction reward shared by all tokens in that turn, and $A_{t,n}^{\mathrm{user}}$ is the corresponding token-level diagnostic advantage. Here, $\mathcal{D}$ denotes the training distribution over users and their interaction trajectories. The $\lambda_{\mathrm{SD}}^{\mathrm{user}}$ controls the trade-off between these two terms.

In this way, CoARS preserves task-specific optimization for RecAgent and UserAgent while enabling their co-evolution within the same interaction loop. Importantly, CoARS trains the RecAgent policy and one shared UserAgent policy, rather than a separate UserAgent for each user. The objective is to optimize the distinct reasoning role of each agent: RecAgent for recommendation reasoning, and UserAgent for evaluation reasoning. Differences across users are captured through historical interaction records and updated memory, which provide personalized context for the shared policies.

\section{Experiments}
\label{Experiments}
We conduct experiments to investigate the effectiveness of the proposed framework from multiple perspectives. In particular, we aim to answer the following research questions: 
\textbf{RQ1:} Can our method improve recommendation accuracy compared with representative agentic recommender baselines? \textbf{RQ2:} Can the evolved UserAgent more accurately simulate user feedback?  
\textbf{RQ3:} How sensitive is the proposed framework to the hyperparameter settings?
\textbf{RQ4:} How much do the key components contribute to the final performance?
\textbf{RQ5:} Which teacher update strategy is more effective for self-distillation, a frozen teacher or an EMA-updated teacher?
\textbf{RQ6:} How does self-distilled credit assignment differ from directly using self-distillation as the optimization objective?


\subsection{Experimental Settings}

\paragraph{Datasets} We conduct experiments on three widely used sequential recommendation benchmarks, including LastFM~\cite{22cantador2011second}, MovieLens~\cite{23harper2015movielens}, and Instruments~\cite{24hou2024bridging}, which cover different recommendation scenarios. We sort each user’s interactions chronologically and construct the data splits based on temporal order~\cite{25zhao2024let,26huang2025recommender}. Due to the large size of the Instruments test set, we randomly sampled 1,000 data points, following the same setting as prior work~\cite{04cai2025agentic} to keep the evaluation scale aligned with LastFM and MovieLens. The overall statistics of the datasets are summarized in Table~\ref{tab:dataset_stats}.

\begin{table}[h]
\centering
\caption{Statistics of the datasets.}
\label{tab:dataset_stats}
\begin{tabular}{lccc}
\toprule
Dataset & \#Users & \#Items & \#Interactions \\
\midrule
LastFM & 2,100 & 18,745 & 92,835 \\
MovieLens & 943 & 1,682 & 100,001 \\
Instruments & 903,330 & 112,136 & 1,512,531 \\
\bottomrule
\end{tabular}
\end{table}

\paragraph{Baselines}

We compare our method with the following representative baselines:

\begin{itemize}[leftmargin=12pt]
    \item {Reflexion~\cite{06shinn2023reflexion}.} A simple Reflexion-based agent that directly predicts the next item from the candidate set based on user history, without introducing the other RecAgent or UserAgent for interactive evaluation.
    \item {AFL}~\cite{04cai2025agentic}. It models the feedback between a RecAgent and UserAgents, where interactions are stored in memory.
    \item {iAgent}~\cite{02xu2025iagent}. A user-side instruction-aware framework that places an LLM agent as a shield between the user and the system. We follow the original setting by simulating user instructions for instruction-aware recommendation.
    \item {RecoWorld}~\cite{05liu2025recoworld}. A simulated environment for ARS that emphasizes reinforcement learning by using the UserAgent's click feedback as a judge signal to provide outcome-level supervision for optimizing the RecAgent.
\end{itemize}

\begin{table}[t]
\centering
\caption{Recommendation performance comparison across datasets. The best result in each dataset column is highlighted in bold, and the second-best result is underlined.}
\label{tab:main_results}
\scriptsize
\setlength{\tabcolsep}{8pt}
\begin{tabular}{llccc}
\toprule
Method & Backbone & LastFM & MovieLens & Instruments \\
\midrule
Reflexion & Qwen3-4B     & 0.0087 & 0.0248 & 0.0158 \\
          & Qwen3-8B     & 0.0096 & 0.0361 & 0.0265 \\
          & GPT-5.4-mini & 0.0169 & 0.0568 & 0.0548 \\
\cmidrule(lr){1-5}
AFL       & Qwen3-4B     & 0.0151 & 0.0404 & 0.0515 \\
          & Qwen3-8B     & 0.0215 & 0.0461 & 0.0624 \\
          & GPT-5.4-mini & 0.0301 & 0.0883 & 0.0811 \\
\cmidrule(lr){1-5}
iAgent    & Qwen3-4B     & 0.0412 & 0.0587 & 0.0890 \\
          & Qwen3-8B     & 0.0648 & 0.0848 & 0.1254 \\
          & GPT-5.4-mini & 0.1583 & 0.1362 & 0.1530 \\
\midrule
RecoWorld & Qwen3-4B     & 0.1248 & 0.1075 & 0.1854 \\
          & Qwen3-8B     & \underline{0.1985} & \underline{0.1724} & \underline{0.2222} \\
\cmidrule(lr){1-5}
CoARS     & Qwen3-4B     & 0.1838 & 0.1505 & 0.2712 \\
          & Qwen3-8B     & \textbf{0.2212} & \textbf{0.2631} & \textbf{0.3470} \\
\bottomrule
\end{tabular}
\end{table}

\paragraph{LLM Backbones} We use three LLM backbones in our experiments: Qwen3-8B\footnote{\url{https://huggingface.co/Qwen/Qwen3-8B}}, Qwen3-4B\footnote{\url{https://huggingface.co/Qwen/Qwen3-4B-Instruct-2507}}, and GPT-5.4-mini\footnote{\url{https://openai.com/}}. Qwen3-8B and Qwen3-4B are used as trainable backbones to instantiate both the RecAgent and the UserAgent, while GPT-5.4 mini is used only for baselines that do not require parameter updating. In our implementation, the overall system is built as an agentic reranker on top of a general recommender. Specifically, we first use SASRec~\cite{01kang2018self} with its best-performing hyperparameters on each dataset to generate a 20-item candidate set for each test instance, consisting of the ground-truth item and the top-19 predicted items. This candidate set is then passed to the agentic framework, which performs multi-turn reasoning, user-side evaluation, and final item selection.

\paragraph{Policy Learning Strategy} We adopt LoRA-based reinforcement fine-tuning~\cite{29hu2022lora}, where interaction reward is combined with self-distillation for optimization. Our implementation uses a frozen-teacher setting, where the teacher remains fixed during training.

\paragraph{Evaluation Metrics} We evaluate the framework from two perspectives. For recommendation performance, we use \textbf{Hit@1} to measure whether the final selected item matches the ground-truth next item. For user simulation performance, we use \textbf{F1} to measure how accurately the UserAgent's decisions align with the preference labels induced by the ground-truth next-item behavior, following prior work on user-agent simulation in recommendation.

\subsection{Recommendation Performance Comparison}

We compare CoARS with representative memory-based methods (\textit{Reflexion}, \textit{AFL}, and \textit{iAgent}) and the RL-based baseline \textit{RecoWorld} across three datasets and multiple LLM backbones to examine whether the proposed co-evolution paradigm can consistently improve recommendation performance. The results are reported in Table~\ref{tab:main_results}, from which we obtain the following findings: Under the Qwen3-8B backbone setting, CoARS consistently achieves the best performance across all datasets and outperforms all competing methods. Even when restricting the comparison to the Qwen3-4B backbone, CoARS still delivers the strongest results, showing that its advantage holds under the same model scale. More importantly, both CoARS and RecoWorld substantially outperform the memory-based methods, indicating that optimizing ARS through parameter-level learning is more effective than relying only on text-level memory reuse.

\begin{tcolorbox}[
colback=gray!8,
colframe=black,
boxrule=0.6pt,
arc=1.5mm,
left=1mm,
right=1mm,
top=0.8mm,
bottom=0.8mm
]
\textbf{Takeaway 1.} CoARS outperforms prior ARS methods across backbone scales, and we highlight the importance of parameter-level learning over text-level memory reuse.
\end{tcolorbox}

\begin{table}[t]
\centering
\caption{User simulation performance comparison across datasets. The best result in each dataset column is highlighted in bold, and the second-best result is underlined.}
\label{tab:main_results_2}
\scriptsize
\setlength{\tabcolsep}{8pt}
\begin{tabular}{llccc}
\toprule
Method & Backbone & LastFM & MovieLens & Instruments \\
\midrule
Reflexion & Qwen3-4B     & 0.0642 & 0.0864 & 0.1043 \\
          & Qwen3-8B     & 0.0896 & 0.1028 & 0.1284 \\
          & GPT-5.4-mini & 0.1182 & 0.1654 & 0.1456 \\
\cmidrule(lr){1-5}
AFL       & Qwen3-4B     & 0.1146 & 0.1445 & 0.1984 \\
          & Qwen3-8B     & 0.1410 & 0.1693 & 0.2384 \\
          & GPT-5.4-mini & 0.2482 & 0.2446 & 0.3484 \\
\cmidrule(lr){1-5}
iAgent    & Qwen3-4B     & 0.1356 & 0.1542 & 0.2642 \\
          & Qwen3-8B     & 0.1572 & 0.1856 & 0.3146 \\
          & GPT-5.4-mini & 0.2653 & \underline{0.2884} & \underline{0.3684} \\
\cmidrule(lr){1-5}
RecoWorld & Qwen3-4B     & 0.1256 & 0.1432 & 0.2254 \\
          & Qwen3-8B     & 0.1513 & 0.1867 & 0.2664 \\
\cmidrule(lr){1-5}
CoARS     & Qwen3-4B     & \underline{0.2164} & 0.1954 & 0.2854 \\
          & Qwen3-8B     & \textbf{0.3145} & \textbf{0.2974} & \textbf{0.3812} \\
\bottomrule
\end{tabular}
\end{table}

\subsection{User Simulation Performance}
We evaluate whether the evolved UserAgent can better simulate user feedback. Specifically, for each test case, we construct a 1:3 candidate set consisting of one ground-truth item and three top-ranked items selected from other traditional recommendation models. The UserAgent is then asked to judge the candidates by producing feedback actions. Following our protocol, \textit{Click} and \textit{Star} are mapped to \textit{accept}, while \textit{Skip} and \textit{Dislike} are mapped to \textit{reject}. In this way, we can directly measure how well different methods simulate user-side decision behavior. The results are reported in Table~\ref{tab:main_results_2}. We observe that CoARS with Qwen3-8B again achieves the best user simulation performance across all datasets. Even under the 4B backbone, CoARS already delivers the strongest result on LastFM. Moreover, although RecoWorld performs strongly in recommendation performance, its advantage is much less evident at the user simulation level. This suggests that optimizing only the recommender side is insufficient for improving the behavioral fidelity of the UserAgent. In contrast, CoARS explicitly trains both agents in a coupled manner, which leads to more effective user-side evolution and better alignment with realistic feedback behavior.

\begin{tcolorbox}[
colback=gray!8,
colframe=black,
boxrule=0.6pt,
arc=1.5mm,
left=1mm,
right=1mm,
top=0.8mm,
bottom=0.8mm
]
\textbf{Takeaway 2.} CoARS not only improves recommendation performance, but also yields a stronger UserAgent, highlighting the importance of bidirectional optimization for agent co-evolution.
\end{tcolorbox}

\begin{figure}[t]
    \centering
    \includegraphics[width=0.98\columnwidth]{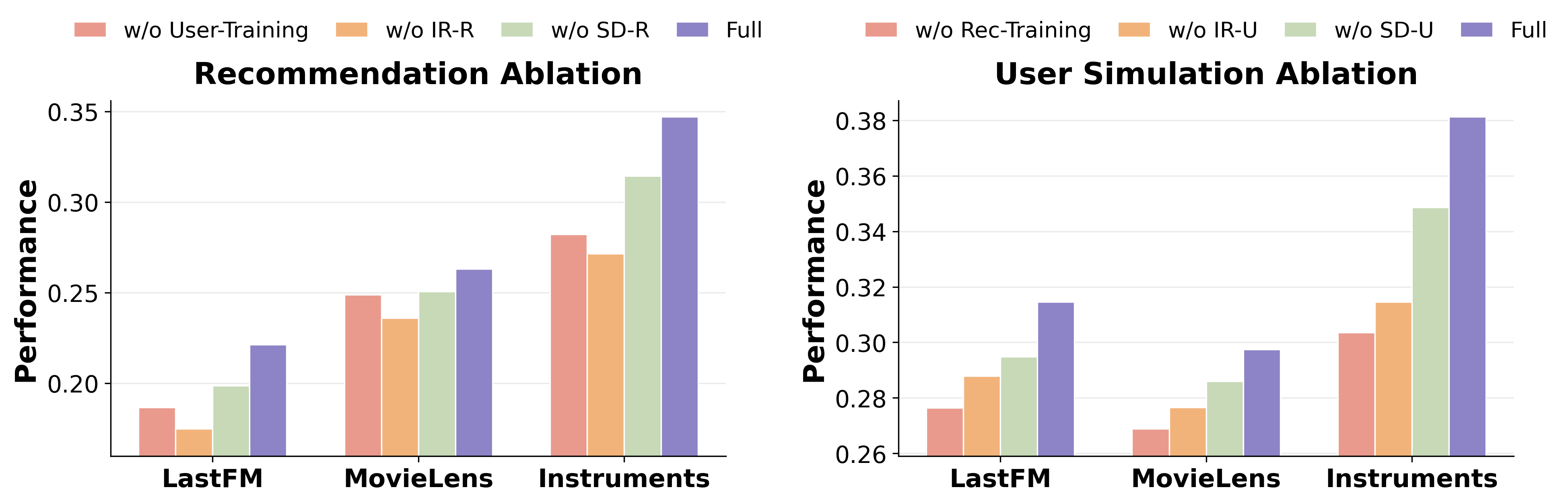}
    \caption{Ablation results of CoARS on the recommender side and the user side.}
    \label{fig:ablation}
\end{figure}

\subsection{The Effect of Hyperparameters}
\begin{figure}[t]
    \centering
    \includegraphics[width=0.95\columnwidth]{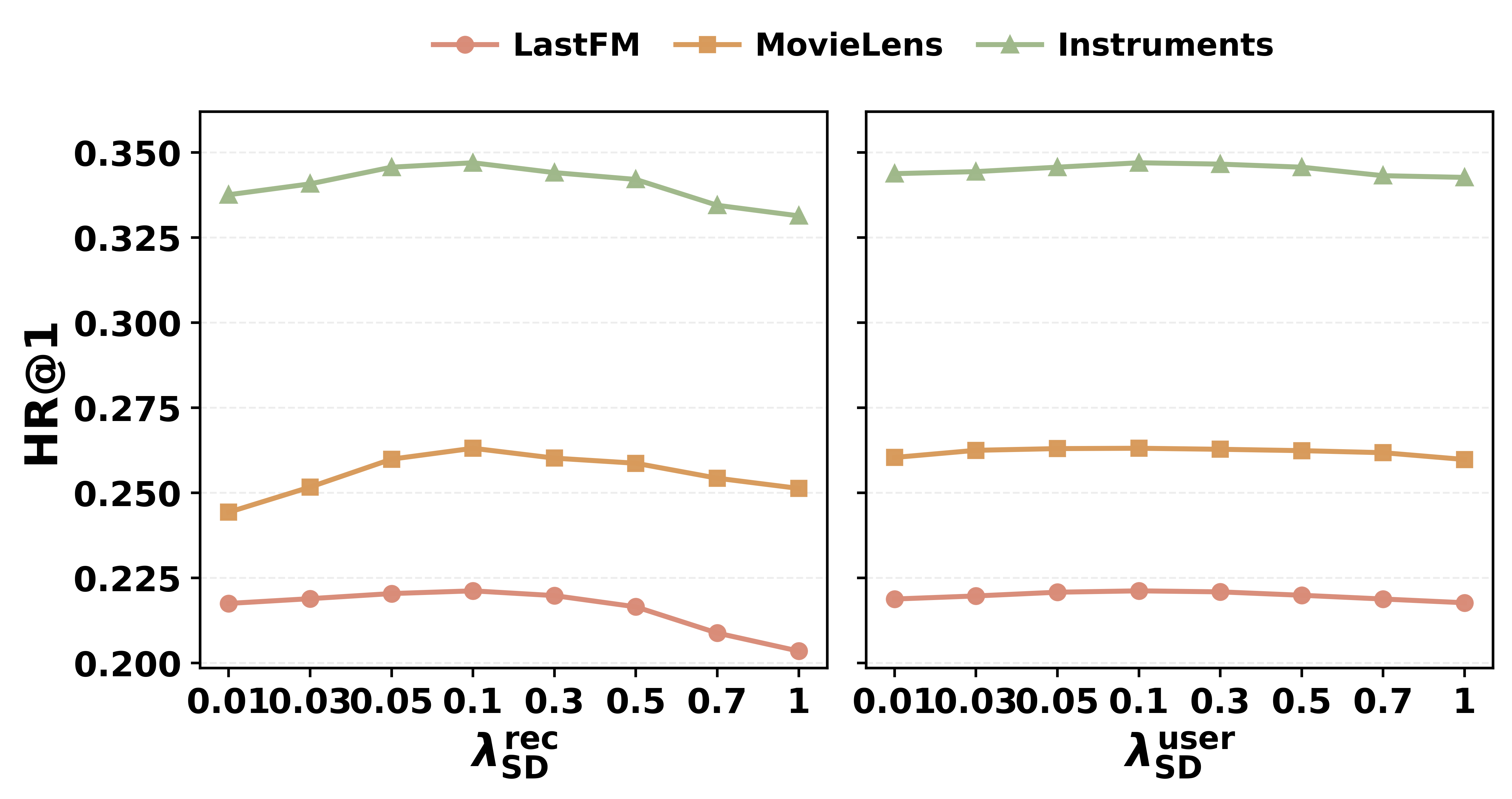}
    \caption{The effect of the hyperparameters.}
    \label{fig:hyper}
\end{figure}

To examine the sensitivity of the two self-distillation coefficients, we separately vary $\lambda_{\mathrm{SD}}^{\mathrm{rec}}$ and $\lambda_{\mathrm{SD}}^{\mathrm{user}}$ over a wide range. Specifically, we vary one coefficient while fixing the other at 0.1. We conduct experiments with Qwen3-8B on three datasets. From Fig.~\ref{fig:hyper}, we observe a consistent inverted-U trend for both coefficients across all three datasets. For both $\lambda_{\mathrm{SD}}^{\mathrm{rec}}$ and $\lambda_{\mathrm{SD}}^{\mathrm{user}}$, the performance steadily improves as the value increases from 0.01 to 0.1, but declines when the value is further enlarged. This result suggests that both coefficients require a moderate setting to properly balance the self-distillation signal in optimization. Moreover, the effect of $\lambda_{\mathrm{SD}}^{\mathrm{rec}}$ is more pronounced, indicating that recommendation-side self-distillation plays a more sensitive role in the overall optimization process. Overall, $\lambda_{\mathrm{SD}}^{\mathrm{rec}}=0.1$ and $\lambda_{\mathrm{SD}}^{\mathrm{user}}=0.1$ give the best and most stable performance in our framework.

\begin{tcolorbox}[
colback=gray!8,
colframe=black,
boxrule=0.5pt,
arc=1.2mm,
left=1mm,
right=1mm,
top=0.6mm,
bottom=0.6mm
]
\textbf{Takeaway 3.} Both $\lambda_{\mathrm{SD}}^{\mathrm{rec}}$ and $\lambda_{\mathrm{SD}}^{\mathrm{user}}$ exhibit a consistent inverted-U effect across datasets, with performance being more sensitive to $\lambda_{\mathrm{SD}}^{\mathrm{rec}}$; setting both coefficients to 0.1 achieves the best overall results.
\end{tcolorbox}

\subsection{Ablation Study}
We conduct ablation studies to examine the contribution of each design in CoARS from both the recommender side and the user side. For RecAgent, we remove the dynamically evolved UserAgent (w/o User-Training) to evaluate the role of co-evolution, and further remove Interaction Reward (w/o IR-R) and Self-Distillation (w/o SD-R) to assess the contribution of reward supervision and token-level credit assignment on the recommender side. Similarly, for UserAgent, we remove the jointly trained RecAgent (w/o Rec-Training) to test the effect of bidirectional evolution, and further remove Interaction Reward (w/o IR-U) and Self-Distillation (w/o SD-U) to analyze their contribution on the user side. As shown in Fig.~\ref{fig:ablation}, every component contributes positively to the final performance. On the recommender side, removing the evolved UserAgent leads to consistent drops on all datasets, showing the importance of co-evolution. Removing Interaction Reward causes the largest degradation, indicating that reward-based supervision is the most critical signal for improving RecAgent. Removing Self-Distillation also consistently hurts performance, confirming the benefit of token-level guidance. A similar pattern is observed on the user side. Once the jointly trained RecAgent is removed, the performance of UserAgent drops substantially, suggesting that user evolution also relies on a strong and adaptive recommender partner. Removing Interaction Reward or Self-Distillation further weakens the results. These findings show that each design in CoARS is effective, and that both agents benefit from bidirectional parameter-level co-evolution.

\begin{tcolorbox}[
colback=gray!8,
colframe=black,
boxrule=0.5pt,
arc=1.2mm,
left=1mm,
right=1mm,
top=0.6mm,
bottom=0.6mm
]
\textbf{Takeaway 4.} Each design in CoARS is effective, and the full model performs best on both recommendation and user simulation, highlighting the importance of bidirectional co-evolution, interaction reward, and self-distilled parameter-level learning.
\end{tcolorbox}

\subsection{The Comparison of Fixed and EMA Teacher Update}
We compare two teacher-mode parameter update strategies for self-distillation credit assignment. The first follows the EMA-teacher design, where an EMA-smoothed teacher is continuously updated together with the student, so that the same underlying policy effectively serves as both teacher and student~\cite{44zhao2026self}. In our implementation, the EMA update rate is set to 0.05. The second strategy instead fixes the teacher model throughout training, so that the student is always distilled toward a stationary teacher. We conduct experiments with Qwen3-8B on three datasets, and the results are shown in Table~\ref{tab:teacher_update_mode}. Our empirical findings consistently show that the fixed-teacher strategy achieves better performance than the EMA-updated teacher. Therefore, we adopt the fixed-teacher mode in our experiments.

\begin{table}[h]
\centering
\caption{Comparison of different teacher-mode parameter update strategies for self-distillation credit assignment. The best result in each column is highlighted in bold.}
\label{tab:teacher_update_mode}
\small
\setlength{\tabcolsep}{10pt}
\begin{tabular}{lccc}
\toprule
Strategy & LastFM & MovieLens & Instruments \\
\midrule
EMA   & 0.2167 & 0.2456 & 0.3278 \\
Fixed & \textbf{0.2212} & \textbf{0.2631} & \textbf{0.3470} \\
\bottomrule
\end{tabular}
\end{table}

\begin{tcolorbox}[
colback=gray!8,
colframe=black,
boxrule=0.5pt,
arc=1.2mm,
left=1mm,
right=1mm,
top=0.6mm,
bottom=0.6mm
]
\textbf{Takeaway 5.} For self-distillation-based credit assignment, a fixed teacher is more effective than an EMA-updated teacher. This suggests that stable, stationary supervision is more beneficial than a moving teacher target.
\end{tcolorbox}

\subsection{The Comparison of Direct Self-Distillation and Our Self-Distilled Credit Assignment}

In our setting, we adopt self-distilled credit assignment instead of directly optimizing self-distillation. To compare these two designs, we conduct experiments with Qwen3-8B on three datasets, as shown in Table~\ref{tab:sd_vs_credit_assignment}. The results show that both designs are effective, while our self-distilled credit assignment consistently performs slightly better. We attribute this to the fact that interaction reward remains the primary learning signal in our framework, while self-distillation mainly provides finer token-level guidance on where stronger updates are needed. Thus, direct KD and token-level credit assignment play similar roles, but the latter is more naturally aligned with the RL objective. Moreover, direct self-distillation requires computing teacher-student divergence over the full vocabulary at every decoding step, whereas our method avoids this extra cost by using token-level rewards, making it more efficient.

\begin{table}[h]
\centering
\caption{Comparison between direct self-distillation and our self-distilled credit assignment. The best result in each column is highlighted in bold.}
\label{tab:sd_vs_credit_assignment}
\small
\setlength{\tabcolsep}{10pt}
\begin{tabular}{lccc}
\toprule
Method & LastFM & MovieLens & Instruments \\
\midrule
Direct-SD   & 0.2198 & 0.2622 & 0.3428 \\
Ours & \textbf{0.2212} & \textbf{0.2631} & \textbf{0.3470} \\
\bottomrule
\end{tabular}
\end{table}

\begin{tcolorbox}[
colback=gray!8,
colframe=black,
boxrule=0.5pt,
arc=1.2mm,
left=1mm,
right=1mm,
top=0.6mm,
bottom=0.6mm
]
\textbf{Takeaway 6.} Both direct self-distillation and self-distilled credit assignment are effective. Our design achieves slightly better performance and higher efficiency by using teacher guidance as token-level reward shaping.
\end{tcolorbox}

\section{Case Study}
\label{casestudy}
We next illustrate how the teacher-mode prompt is constructed from the student-mode interaction. Specifically, we first show the concrete inputs and outputs of RecAgent and UserAgent in student mode. We then use these outputs, together with the diagnostic label, to build the reference block. This block is subsequently inserted into the unified prompt template to form the teacher-mode input.

\paragraph{Student-Mode Inputs and Outputs}
\label{app:student_mode_io}

\begin{tcolorbox}[
    title={RecAgent: Student-Mode Input and Output},
    colback=white,
    colframe=black,
    breakable,
    sharp corners
]
\ttfamily\small

\InputTag\\
\begingroup
\color{inputcolor}
You are a movie recommendation system. Refine the user's watching history to predict the most likely movie they will watch next from a selection of candidates. However, the user might feel that the movie you recommended is not their top choice from the list of candidates. Based on the above information, select the best movie again from the candidate list.

\color{inputcolor}
\textcolor{inputcolor}{Some useful tips:} \\
\color{inputcolor}
1. You need to first give the reasons, and then provide the recommended movie.\\
2. The movie you recommend must be in the candidate list.\\
You must follow this output format:\\
\color{inputcolor}
Reason: <your reason example>\\
Item: <item example>

\Var{This user has watched Austin Powers: International Man of Mystery, Grosse Pointe Blank, Grumpier Old Men, Eddie, Romy and Michele's High School Reunion, Don't Be a Menace to South Central While Drinking Your Juice in the Hood, Addicted to Love, Bulletproof, Bogus, Toy Story in the previous.}\\
Given the following movies: \Var{Mrs. Dalloway, Reluctant Debutante, The, Heaven \& Earth, Muppet Treasure Island, Jane Eyre, Independence Day (ID4), Associate, The, Basketball Diaries, Ghost in the Shell (Kokaku kidotai), Walking Dead, The, Apt Pupil, Ripe, War at Home, The, Nothing Personal, Maltese Falcon, The, Getting Even with Dad, Pather Panchali, Living in Oblivion, Sleepover, you should recommend one movie for this user to watch next.}\\
Here are your previous recommendation records and feedback (it may be empty at the beginning):\\
\Var{In round 1, the movie you recommended is Rock.\\
The reason you gave for the recommendation is: The user has a history of watching movies that are comedic, quirky, and often have a mix of humor and heart. Films like Austin Powers: International Man of Mystery, Grosse Pointe Blank, and Romy and Michele's High School Reunion suggest a preference for light-hearted, character-driven stories with a touch of absurdity. Among the candidate list, Rock fits this pattern as it is a comedy that balances humor with a heartfelt message, making it a suitable follow-up to the user's previous choices.}\\
\Var{The reason the user provided for not considering this to be the best recommendation is: The user's history shows a preference for comedic, quirky, and heartwarming films, not intense psychological dramas. Rock leans more into drama and tension, which doesn't align with the user's past preferences.}

Based on the above information, select the best movie again from the candidate list.
\par
\endgroup

\vspace{0.5em}

\OutputTag\\
\begingroup
\color{outputcolor}
Reason: The user has a history of watching comedic, quirky, and heartwarming films such as *Austin Powers: International Man of Mystery*, *Grosse Pointe Blank*, and *Romy 
and Michele's High School Reunion*. These films suggest a preference for 
\color{outputcolor}light-hearted, character-driven stories with humor and a touch of absurdity. Among the candidate list, *Sleepover* fits this pattern as it is a comedy that balances humor with a heartfelt message, making it a suitable follow-up to the user's previous choices.\\
Item: Sleepover
\par
\endgroup

\end{tcolorbox}

\begin{tcolorbox}[
    title={UserAgent: Student-Mode Input and Output},
    colback=white,
    colframe=black,
    breakable,
    sharp corners
]
\ttfamily\small

\InputTag\\
\begingroup
\color{inputcolor}
The list of candidate movies is: \Var{Mrs. Dalloway, Reluctant Debutante, The, Heaven \& Earth, Muppet Treasure Island, Jane Eyre, Independence Day (ID4), Associate, The, Basketball Diaries, Ghost in the Shell (Kokaku kidotai), Walking Dead, The, Apt Pupil, Ripe, War at Home, The, Nothing Personal, Maltese Falcon, The, Rock, Getting Even with Dad, Pather Panchali, Living in Oblivion, Sleepover}.\\
You can focus on considering these movies: \Var{Rock, Independence Day (ID4), Muppet Treasure Island, Associate, The, Jane Eyre, Apt Pupil, Ghost in the Shell (Kokaku kidotai), Maltese Falcon, The, Getting Even with Dad, Basketball Diaries}.\\
Here are the movies previously recommended by the recommendation system and the reasons for these recommendations, along with your reasons for thinking that the recommended movies were not the best choices:\\
\Var{In round 1, the recommended movie is Rock.}\\
\Var{The reason given by the recommendation system is: The user has a history of watching movies that are comedic, quirky, and often have a mix of humor and heart. Films like Austin Powers: International Man of Mystery, Grosse Pointe Blank, and Romy and Michele's High School Reunion suggest a preference for light-hearted, character-driven stories with a touch of absurdity. Among the candidate list, Rock fits this pattern as it is a comedy that balances humor with a heartfelt message, making it a suitable follow-up to the user's previous choices.}\\
\Var{The reason you provided for not considering this the best recommendation is: The user's history shows a preference for comedic, quirky, and heartwarming films, not intense psychological dramas. Rock leans more into drama and tension, which doesn't align with the user's past preferences.}\\

Now, the new movie recommended by the recommendation system is: \Var{Sleepover}.\\
The recommendation system provides the following reason: \Var{The user has a}
\Var{history of watching comedic, quirky, and heartwarming films such as Austin Powers: International Man of Mystery, Grosse Pointe Blank, and Romy and Michele's High School Reunion. These films suggest a preference for light-hearted, character-driven stories with humor and a touch of absurdity. Among the candidate list, Sleepover fits this pattern as it is a comedy that balances humor with a heartfelt message, making it a suitable follow-up to the user's previous choices.}\\
\color{inputcolor}
Another user who has interacted with this newly recommended movie says they like it because: \Var{Sleepover might match my interests because I enjoy movies that blend humor with heartfelt moments, similar to That Darn Cat! and Sleeper, which also feature quirky characters and light-hearted storytelling. Additionally, Sleepover could appeal to my appreciation for character-driven narratives, much like The Quiet Man and Nobody's Fool, which focus on personal growth and emotional depth.}

Based on the above information, please determine if the newly recommended movie is the most preferred one on the candidate list for you.
\par
\endgroup

\vspace{0.5em}

\OutputTag\\
\begingroup
\color{outputcolor}
Reason: The user's history shows a preference for comedic, quirky, and heartwarming films, not intense or dramatic stories. Sleepover fits this pattern with its comedic and character-driven approach, aligning well with the user's past preferences.\\
Decision: click\\
Strength: 0.91
\par
\endgroup

\end{tcolorbox}

\paragraph{Constructing Reference Trajectory and Forming Teacher Mode}
\label{app:demo_teacher_construction}

Based on the student-mode outputs above, we compare the generated interaction with the target item and the expected user response to derive a diagnostic label. The resulting diagnosed trajectory is then rewritten into a reference trajectory, which is inserted into the teacher-mode prompt as a reference block for both the RecAgent and the UserAgent, yielding their respective teacher-mode inputs for alignment.

\begin{tcolorbox}[
    title={Constructed Reference and Teacher-Mode Output},
    colback=white,
    colframe=black,
    breakable,
    sharp corners
]
\ttfamily\small

\DemoTag\\
\begingroup
\color{democolor}
Current recommendation: \Var{Sleepover}\\
Current recommendation reason: \Var{The user has a history of watching comedic, quirky, and heartwarming films such as *Austin Powers: International Man of Mystery*, *Grosse Pointe Blank*, and *Romy and Michele's High School Reunion*. These films suggest a preference for light-hearted, character-driven stories with humor
and a touch of absurdity. Among the candidate list, *Sleepover* fits this } 
\textcolor{varcolor}{pattern as it is a comedy that balances humor with a heartfelt message, making it a suitable follow-up to the user's previous choices.}\\
\color{democolor}{User's actual response:} \Var{action=click, binary\_decision=yes, reason=The user's history shows a preference for comedic, quirky, and heartwarming films, not intense or dramatic stories. "Sleepover" fits this pattern with its comedic and character-driven approach, aligning well with the user's past preferences.}\\
Ground-truth target item: \Var{Rock}\\
Reference reasoning: \Var{The user's history indicates a strong preference for comedic, quirky, and heartwarming films with a mix of humor and character-driven storytelling. Movies like *Austin Powers: International Man of Mystery*, *Grosse Pointe Blank*, and *Romy and Michele's High School Reunion* all share these traits. While *Rock* was initially recommended, the user expressed that it leaned too heavily into drama and tension, which doesn't match their preference for lighthearted and humorous content. Given the limited candidate list and the need to align with the user's stated preferences, there are no other movies in the list that better match their taste for comedy and character-driven stories. Therefore, *Rock* remains the most suitable choice despite the user's initial hesitation, as it is the only option that fits the comedic and character-driven pattern they have shown in their history.}\\
Reference user response for this case: \Var{skip\_or\_dislike (example: dislike), binary\_decision=no}\\
Reference reason: \Var{Use skip or dislike when recommendation does not match target preference.}\\
Diagnosis: \Var{rec\_agent=wrong, user\_agent=wrong, outcome=rec\_wrong\_user\_accepted}
\par
\endgroup

\end{tcolorbox}

From the above content, we can draw several insights: First, the teacher-mode signal is tightly aligned with the actual failure mode of the current trajectory, rather than providing a generic correction. Second, the diagnostic label explicitly disentangles whether the error comes from the RecAgent, the UserAgent, or both, so that the constructed reference trajectory serves as targeted supervision rather than undifferentiated replay. Third, the reference trajectory preserves both the recommendation-side reasoning and the user-side response.


\section{Related Work}
\label{related}
Recent work on human-AI collaboration has increasingly emphasized that intelligent systems should not merely provide outputs, but also support iterative preference elicitation and preserve user agency throughout the interaction process~\cite{48wu2025sharedagency}. Against this shift, early agentic recommender systems mainly evolve through memory update under the Reflexion-style paradigm~\cite{27wang2025user,28wang2024recmind,42wang2025id}, where historical interactions are appended to memory and reused as prompt context in later turns~\cite{03chen2026memrec,13zhang2024agentcf,14liu2025agentcf++,15xia2025multi,17li2026recnet}. This design improves cross-turn consistency and allows agents to recall prior successes, failures, and preference cues, but its effect remains largely at the prompt level, since historical interactions are replayed as text rather than converted into direct parameter-level supervision.

More recent work has introduced reinforcement learning~\cite{12shao2024deepseekmath,08guo2025deepseek} to optimize ARS at the parameter level. These methods typically assign rewards according to recommendation success, using either human-designed reward functions or agent-as-judge signals~\cite{05liu2025recoworld,09nguyen2026amem4rec}. However, existing RL-based ARS still mostly rely on sparse outcome-level supervision and one-sided optimization, where only one agent is trained while the other is treated as fixed. As a result, they provide limited guidance on which reasoning steps or feedback behaviors should be encouraged. Recently, the self-distillation paradigm~\cite{18shenfeld2026self,19hubotter2026reinforcement,20zhao2026self} has been used to provide finer-grained token-level supervision by comparing a student model with a teacher version of itself under richer guidance. Such methods show that privileged reference trajectories can provide more informative credit assignment than scalar rewards alone. Our work connects these lines by showing that ARS interaction histories naturally support both bidirectional interaction reward and self-distilled token-level credit assignment, enabling the co-evolution of RecAgent and UserAgent.

\section{Conclusion}
\label{conclusion}
In this paper, we investigate how historical interactions can be more fully exploited to drive the co-evolution of agents in agentic recommender systems. We propose CoARS, a self-distilled reinforcement learning framework for co-evolving agentic recommender systems. Unlike existing methods that mainly reuse historical interactions as external memory or optimize only one agent with sparse outcome-level rewards, CoARS transforms historical interactions into richer learning signals for both agents through interaction reward and self-distilled credit assignment. Experiments on multiple benchmarks show that CoARS consistently improves recommendation accuracy and user alignment, demonstrating the effectiveness of leveraging historical interactions as a stronger training signal for agent co-evolution.

Looking ahead, we plan to further explore interaction-driven supervision as a foundation for building more robust agentic recommender systems, with a particular focus on enhancing personalization quality and improving the robustness of agent behavior under noisy or uncertain feedback. We are also interested in improving the efficiency and scalability of interaction-based optimization to support faster and more responsive multi-turn interactions. In addition, an important future direction is to incorporate stronger safety considerations into the co-evolution process, such as making agent behaviors more reliable, reducing harmful or biased interaction patterns, and improving robustness against misleading or adversarial feedback. We hope this line of research can further advance agentic recommender systems from simple memory reuse toward parameter adaptation through interaction.

\bibliography{CoARS}

@inproceedings{01kang2018self,
  title={Self-attentive sequential recommendation},
  author={Kang, Wang-Cheng and McAuley, Julian},
  booktitle={2018 IEEE international conference on data mining (ICDM)},
  pages={197--206},
  year={2018},
  organization={IEEE}
}

@inproceedings{02xu2025iagent,
  title={iagent: Llm agent as a shield between user and recommender systems},
  author={Xu, Wujiang and Shi, Yunxiao and Liang, Zujie and Ning, Xuying and Mei, Kai and Wang, Kun and Zhu, Xi and Xu, Min and Zhang, Yongfeng},
  booktitle={Findings of the Association for Computational Linguistics: ACL 2025},
  pages={18056--18084},
  year={2025}
}

@article{03chen2026memrec,
  title={MemRec: Collaborative Memory-Augmented Agentic Recommender System},
  author={Chen, Weixin and Zhao, Yuhan and Huang, Jingyuan and Ye, Zihe and Ju, Clark Mingxuan and Zhao, Tong and Shah, Neil and Chen, Li and Zhang, Yongfeng},
  journal={arXiv preprint arXiv:2601.08816},
  year={2026}
}

@inproceedings{04cai2025agentic,
  title={Agentic feedback loop modeling improves recommendation and user simulation},
  author={Cai, Shihao and Zhang, Jizhi and Bao, Keqin and Gao, Chongming and Wang, Qifan and Feng, Fuli and He, Xiangnan},
  booktitle={Proceedings of the 48th International ACM SIGIR conference on Research and Development in Information Retrieval},
  pages={2235--2244},
  year={2025}
}

@article{05liu2025recoworld,
  title={Recoworld: Building simulated environments for agentic recommender systems},
  author={Liu, Fei and Lin, Xinyu and Yu, Hanchao and Wu, Mingyuan and Wang, Jianyu and Zhang, Qiang and Zhao, Zhuokai and Xia, Yinglong and Zhang, Yao and Li, Weiwei and others},
  journal={arXiv preprint arXiv:2509.10397},
  year={2025}
}

@article{06shinn2023reflexion,
  title={Reflexion: Language agents with verbal reinforcement learning},
  author={Shinn, Noah and Cassano, Federico and Gopinath, Ashwin and Narasimhan, Karthik and Yao, Shunyu},
  journal={Advances in neural information processing systems},
  volume={36},
  pages={8634--8652},
  year={2023}
}

@article{07you2026agent,
  title={Agent-as-a-Judge},
  author={You, Runyang and Cai, Hongru and Zhang, Caiqi and Xu, Qiancheng and Liu, Meng and Yu, Tiezheng and Li, Yongqi and Li, Wenjie},
  journal={arXiv preprint arXiv:2601.05111},
  year={2026}
}

@article{08guo2025deepseek,
  title={DeepSeek-R1 incentivizes reasoning in LLMs through reinforcement learning},
  author={Guo, Daya and Yang, Dejian and Zhang, Haowei and Song, Junxiao and Wang, Peiyi and Zhu, Qihao and Xu, Runxin and Zhang, Ruoyu and Ma, Shirong and Bi, Xiao and others},
  journal={Nature},
  volume={645},
  number={8081},
  pages={633--638},
  year={2025},
  publisher={Nature Publishing Group UK London}
}

@article{09nguyen2026amem4rec,
  title={AMEM4Rec: Leveraging Cross-User Similarity for Memory Evolution in Agentic LLM Recommenders},
  author={Nguyen, Minh-Duc and Kieu, Hai-Dang and Le, Dung D},
  journal={arXiv preprint arXiv:2602.08837},
  year={2026}
}

@inproceedings{10he2020lightgcn,
  title={Lightgcn: Simplifying and powering graph convolution network for recommendation},
  author={He, Xiangnan and Deng, Kuan and Wang, Xiang and Li, Yan and Zhang, Yongdong and Wang, Meng},
  booktitle={Proceedings of the 43rd International ACM SIGIR conference on research and development in Information Retrieval},
  pages={639--648},
  year={2020}
}

@inproceedings{11zhang2024generative,
  title={On generative agents in recommendation},
  author={Zhang, An and Chen, Yuxin and Sheng, Leheng and Wang, Xiang and Chua, Tat-Seng},
  booktitle={Proceedings of the 47th international ACM SIGIR conference on research and development in Information Retrieval},
  pages={1807--1817},
  year={2024}
}

@article{12shao2024deepseekmath,
  title={Deepseekmath: Pushing the limits of mathematical reasoning in open language models},
  author={Shao, Zhihong and Wang, Peiyi and Zhu, Qihao and Xu, Runxin and Song, Junxiao and Bi, Xiao and Zhang, Haowei and Zhang, Mingchuan and Li, YK and Wu, Yang and others},
  journal={arXiv preprint arXiv:2402.03300},
  year={2024}
}

@inproceedings{13zhang2024agentcf,
  title={Agentcf: Collaborative learning with autonomous language agents for recommender systems},
  author={Zhang, Junjie and Hou, Yupeng and Xie, Ruobing and Sun, Wenqi and McAuley, Julian and Zhao, Wayne Xin and Lin, Leyu and Wen, Ji-Rong},
  booktitle={Proceedings of the ACM Web Conference 2024},
  pages={3679--3689},
  year={2024}
}

@inproceedings{14liu2025agentcf++,
  title={AgentCF++: Memory-enhanced LLM-based Agents for Popularity-aware Cross-domain Recommendations},
  author={Liu, Jiahao and Gu, Shengkang and Li, Dongsheng and Zhang, Guangping and Han, Mingzhe and Gu, Hansu and Zhang, Peng and Lu, Tun and Shang, Li and Gu, Ning},
  booktitle={Proceedings of the 48th International ACM SIGIR Conference on Research and Development in Information Retrieval},
  pages={2566--2571},
  year={2025}
}

@article{15xia2025multi,
  title={Multi-Agent Collaborative Filtering: Orchestrating Users and Items for Agentic Recommendations},
  author={Xia, Yu and Kim, Sungchul and Yu, Tong and Rossi, Ryan A and McAuley, Julian},
  journal={arXiv preprint arXiv:2511.18413},
  year={2025}
}

@article{17li2026recnet,
  title={RecNet: Self-Evolving Preference Propagation for Agentic Recommender Systems},
  author={Li, Bingqian and Wang, Xiaolei and Li, Junyi and Li, Weitao and Zhang, Long and Chen, Sheng and Zhao, Wayne Xin and Wen, Ji-Rong},
  journal={arXiv preprint arXiv:2601.21609},
  year={2026}
}

@article{18shenfeld2026self,
  title={Self-Distillation Enables Continual Learning},
  author={Shenfeld, Idan and Damani, Mehul and H{\"u}botter, Jonas and Agrawal, Pulkit},
  journal={arXiv preprint arXiv:2601.19897},
  year={2026}
}

@article{19hubotter2026reinforcement,
  title={Reinforcement Learning via Self-Distillation},
  author={H{\"u}botter, Jonas and L{\"u}beck, Frederike and Behric, Lejs and Baumann, Anton and Bagatella, Marco and Marta, Daniel and Hakimi, Ido and Shenfeld, Idan and Buening, Thomas Kleine and Guestrin, Carlos and others},
  journal={arXiv preprint arXiv:2601.20802},
  year={2026}
}

@article{20zhao2026self,
  title={Self-Distilled Reasoner: On-Policy Self-Distillation for Large Language Models},
  author={Zhao, Siyan and Xie, Zhihui and Liu, Mengchen and Huang, Jing and Pang, Guan and Chen, Feiyu and Grover, Aditya},
  journal={arXiv preprint arXiv:2601.18734},
  year={2026}
}

@inproceedings{21wang2023efficient,
  title={Efficient bi-level optimization for recommendation denoising},
  author={Wang, Zongwei and Gao, Min and Li, Wentao and Yu, Junliang and Guo, Linxin and Yin, Hongzhi},
  booktitle={Proceedings of the 29th ACM SIGKDD conference on knowledge discovery and data mining},
  pages={2502--2511},
  year={2023}
}

@inproceedings{22cantador2011second,
  title={Second workshop on information heterogeneity and fusion in recommender systems (HetRec2011)},
  author={Cantador, Iv{\'a}n and Brusilovsky, Peter and Kuflik, Tsvi},
  booktitle={Proceedings of the fifth ACM conference on Recommender systems},
  pages={387--388},
  year={2011}
}

@article{23harper2015movielens,
  title={The movielens datasets: History and context},
  author={Harper, F Maxwell and Konstan, Joseph A},
  journal={Acm transactions on interactive intelligent systems (tiis)},
  volume={5},
  number={4},
  pages={1--19},
  year={2015},
  publisher={Acm New York, NY, USA}
}

@article{24hou2024bridging,
  title={Bridging language and items for retrieval and recommendation},
  author={Hou, Yupeng and Li, Jiacheng and He, Zhankui and Yan, An and Chen, Xiusi and McAuley, Julian},
  journal={arXiv preprint arXiv:2403.03952},
  year={2024}
}

@inproceedings{25zhao2024let,
  title={Let me do it for you: Towards llm empowered recommendation via tool learning},
  author={Zhao, Yuyue and Wu, Jiancan and Wang, Xiang and Tang, Wei and Wang, Dingxian and De Rijke, Maarten},
  booktitle={Proceedings of the 47th International ACM SIGIR Conference on Research and Development in Information Retrieval},
  pages={1796--1806},
  year={2024}
}

@article{26huang2025recommender,
  title={Recommender ai agent: Integrating large language models for interactive recommendations},
  author={Huang, Xu and Lian, Jianxun and Lei, Yuxuan and Yao, Jing and Lian, Defu and Xie, Xing},
  journal={ACM Transactions on Information Systems},
  volume={43},
  number={4},
  pages={1--33},
  year={2025},
  publisher={ACM New York, NY}
}

@article{27wang2025user,
  title={User behavior simulation with large language model-based agents},
  author={Wang, Lei and Zhang, Jingsen and Yang, Hao and Chen, Zhi-Yuan and Tang, Jiakai and Zhang, Zeyu and Chen, Xu and Lin, Yankai and Sun, Hao and Song, Ruihua and others},
  journal={ACM Transactions on Information Systems},
  volume={43},
  number={2},
  pages={1--37},
  year={2025},
  publisher={ACM New York, NY}
}

@inproceedings{28wang2024recmind,
  title={Recmind: Large language model powered agent for recommendation},
  author={Wang, Yancheng and Jiang, Ziyan and Chen, Zheng and Yang, Fan and Zhou, Yingxue and Cho, Eunah and Fan, Xing and Lu, Yanbin and Huang, Xiaojiang and Yang, Yingzhen},
  booktitle={Findings of the Association for Computational Linguistics: NAACL 2024},
  pages={4351--4364},
  year={2024}
}

@article{29hu2022lora,
  title={Lora: Low-rank adaptation of large language models.},
  author={Hu, Edward J and Shen, Yelong and Wallis, Phillip and Allen-Zhu, Zeyuan and Li, Yuanzhi and Wang, Shean and Wang, Liang and Chen, Weizhu and others},
  journal={Iclr},
  volume={1},
  number={2},
  pages={3},
  year={2022}
}

@article{32yu2025dapo,
  title={Dapo: An open-source llm reinforcement learning system at scale},
  author={Yu, Qiying and Zhang, Zheng and Zhu, Ruofei and Yuan, Yufeng and Zuo, Xiaochen and Yue, Yu and Dai, Weinan and Fan, Tiantian and Liu, Gaohong and Liu, Lingjun and others},
  journal={arXiv preprint arXiv:2503.14476},
  year={2025}
}

@article{33schulman2017proximal,
  title={Proximal policy optimization algorithms},
  author={Schulman, John and Wolski, Filip and Dhariwal, Prafulla and Radford, Alec and Klimov, Oleg},
  journal={arXiv preprint arXiv:1707.06347},
  year={2017}
}

@article{34tran2026entropy,
  title={Entropy Guided Diversification and Preference Elicitation in Agentic Recommendation Systems},
  author={Tran, Dat and Li, Yongce and Clay, Hannah and Golrezaei, Negin and Beygi, Sajjad and Saberi, Amin},
  journal={arXiv preprint arXiv:2603.11399},
  year={2026}
}

@article{35song2026expanding,
  title={Expanding the Capabilities of Reinforcement Learning via Text Feedback},
  author={Song, Yuda and Chen, Lili and Tajwar, Fahim and Munos, Remi and Pathak, Deepak and Bagnell, J Andrew and Singh, Aarti and Zanette, Andrea},
  journal={arXiv preprint arXiv:2602.02482},
  year={2026}
}

@inproceedings{36hou2025treerl,
  title={Treerl: Llm reinforcement learning with on-policy tree search},
  author={Hou, Zhenyu and Hu, Ziniu and Li, Yujiang and Lu, Rui and Tang, Jie and Dong, Yuxiao},
  booktitle={Proceedings of the 63rd Annual Meeting of the Association for Computational Linguistics (Volume 1: Long Papers)},
  pages={12355--12369},
  year={2025}
}

@article{37wen2025reinforcement,
  title={Reinforcement learning with verifiable rewards implicitly incentivizes correct reasoning in base llms},
  author={Wen, Xumeng and Liu, Zihan and Zheng, Shun and Ye, Shengyu and Wu, Zhirong and Wang, Yang and Xu, Zhijian and Liang, Xiao and Li, Junjie and Miao, Ziming and others},
  journal={arXiv preprint arXiv:2506.14245},
  year={2025}
}

@inproceedings{38zhong2024memorybank,
  title={Memorybank: Enhancing large language models with long-term memory},
  author={Zhong, Wanjun and Guo, Lianghong and Gao, Qiqi and Ye, He and Wang, Yanlin},
  booktitle={Proceedings of the AAAI conference on artificial intelligence},
  volume={38},
  number={17},
  pages={19724--19731},
  year={2024}
}

@inproceedings{39bao2023tallrec,
  title={Tallrec: An effective and efficient tuning framework to align large language model with recommendation},
  author={Bao, Keqin and Zhang, Jizhi and Zhang, Yang and Wang, Wenjie and Feng, Fuli and He, Xiangnan},
  booktitle={Proceedings of the 17th ACM conference on recommender systems},
  pages={1007--1014},
  year={2023}
}

@inproceedings{40wang2024macrec,
  title={Macrec: A multi-agent collaboration framework for recommendation},
  author={Wang, Zhefan and Yu, Yuanqing and Zheng, Wendi and Ma, Weizhi and Zhang, Min},
  booktitle={Proceedings of the 47th International ACM SIGIR Conference on Research and Development in Information Retrieval},
  pages={2760--2764},
  year={2024}
}

@article{41wang2025ruleagent,
  title={RuleAgent: Discovering Rules for Recommendation Denoising with Autonomous Language Agents},
  author={Wang, Zongwei and Gao, Min and Yu, Junliang and Hou, Yupeng and Sadiq, Shazia and Yin, Hongzhi},
  journal={arXiv preprint arXiv:2503.23374},
  year={2025}
}

@inproceedings{42wang2025id,
  title={Id-free not risk-free: Llm-powered agents unveil risks in id-free recommender systems},
  author={Wang, Zongwei and Gao, Min and Yu, Junliang and Gao, Xinyi and Nguyen, Quoc Viet Hung and Sadiq, Shazia and Yin, Hongzhi},
  booktitle={Proceedings of the 48th International ACM SIGIR Conference on Research and Development in Information Retrieval},
  pages={1902--1911},
  year={2025}
}

@article{43yang2026self,
  title={Self-Distilled RLVR},
  author={Yang, Chenxu and Qin, Chuanyu and Si, Qingyi and Chen, Minghui and Gu, Naibin and Yao, Dingyu and Lin, Zheng and Wang, Weiping and Wang, Jiaqi and Duan, Nan},
  journal={arXiv preprint arXiv:2604.03128},
  year={2026}
}

@article{44zhao2026self,
  title={Self-Distilled Reasoner: On-Policy Self-Distillation for Large Language Models},
  author={Zhao, Siyan and Xie, Zhihui and Liu, Mengchen and Huang, Jing and Pang, Guan and Chen, Feiyu and Grover, Aditya},
  journal={arXiv preprint arXiv:2601.18734},
  year={2026}
}

@article{45xu2025mem,
  title={A-mem: Agentic memory for llm agents},
  author={Xu, Wujiang and Liang, Zujie and Mei, Kai and Gao, Hang and Tan, Juntao and Zhang, Yongfeng},
  journal={arXiv preprint arXiv:2502.12110},
  year={2025}
}

@article{46lee2023supervised,
  title={Supervised pretraining can learn in-context reinforcement learning},
  author={Lee, Jonathan and Xie, Annie and Pacchiano, Aldo and Chandak, Yash and Finn, Chelsea and Nachum, Ofir and Brunskill, Emma},
  journal={Advances in Neural Information Processing Systems},
  volume={36},
  pages={43057--43083},
  year={2023}
}

@article{47song2025reward,
  title={Reward is enough: Llms are in-context reinforcement learners},
  author={Song, Kefan and Moeini, Amir and Wang, Peng and Gong, Lei and Chandra, Rohan and Zhang, Shangtong and Qi, Yanjun},
  journal={arXiv preprint arXiv:2506.06303},
  year={2025}
}

@inproceedings{48wu2025sharedagency,
  title={Negotiating the Shared Agency between Humans \& AI in the Recommender System},
  author={Wu, Mengke and Liu, Weizi and Wang, Yanyun and Yao, Mike},
  booktitle={Extended Abstracts of the CHI Conference on Human Factors in Computing Systems (CHI EA '25)},
  year={2025},
  doi={10.1145/3706599.3719900}
}
\bibliographystyle{IEEEtran}












\begin{IEEEbiography}[{\includegraphics[width=1in,height=1.25in,clip,]{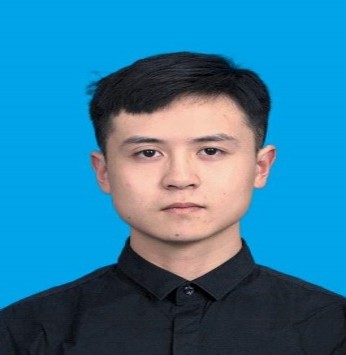}}]{Zongwei Wang}
is currently pursuing his Ph.D. at Chongqing University. His research has been published in top data mining conferences and journals such as KDD, SIGIR, WWW, WSDM, CIKM, and TIST. He has also delivered a tutorial at the top-tier conference WSDM.
\end{IEEEbiography}

\vspace{0.1pt}
\begin{IEEEbiography}[{\includegraphics[width=1in,height=1.1in,clip,]{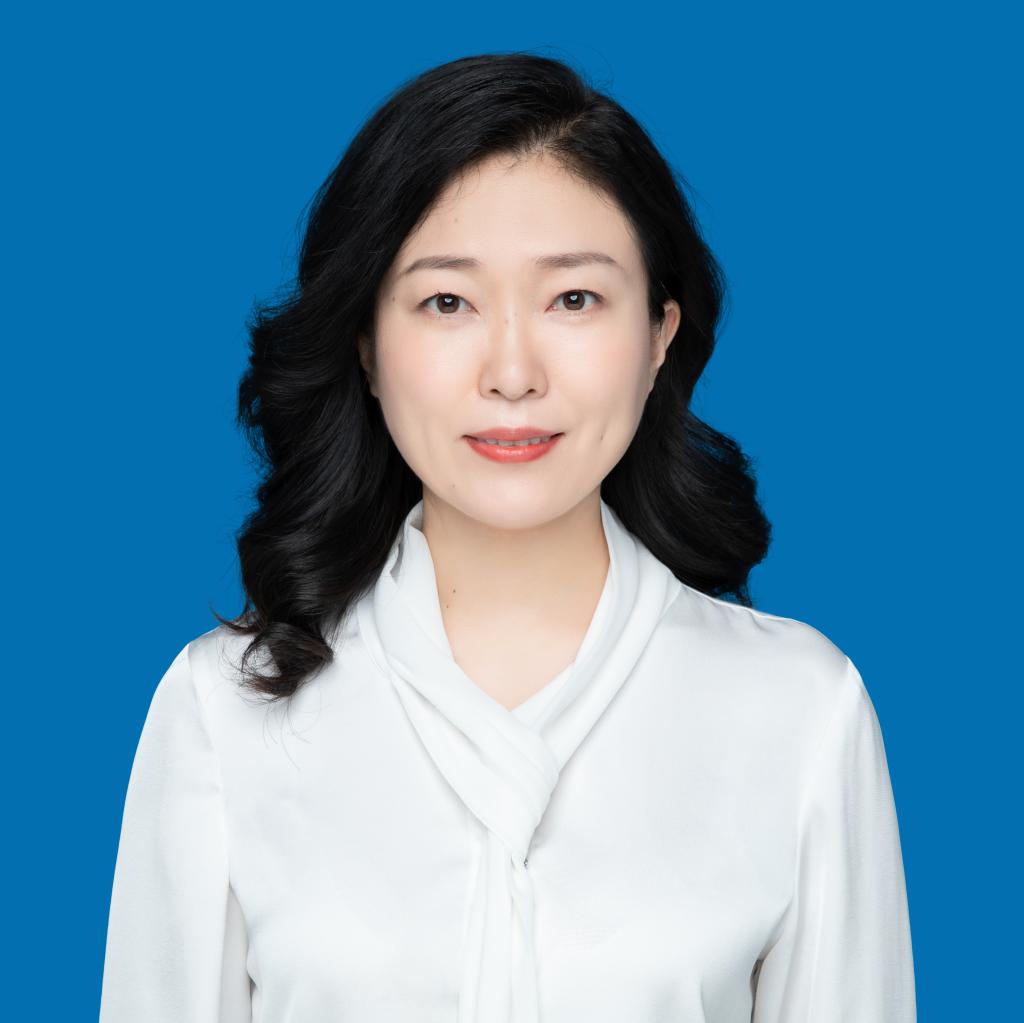}}]{Min Gao}
works as a full professor at Chongqing University, China. Her research areas include recommender systems, anomaly detection, and social media mining. She has published 100+ papers on top data mining conferences and journals such as KDD, SIGIR, WWW, VLDB, WSDM, CIKM, ICDE, TKDD, TIST and TCSS.
\end{IEEEbiography}

\vspace{0.1pt}
\begin{IEEEbiography}[{\includegraphics[width=1in,height=1.25in,clip,keepaspectratio]{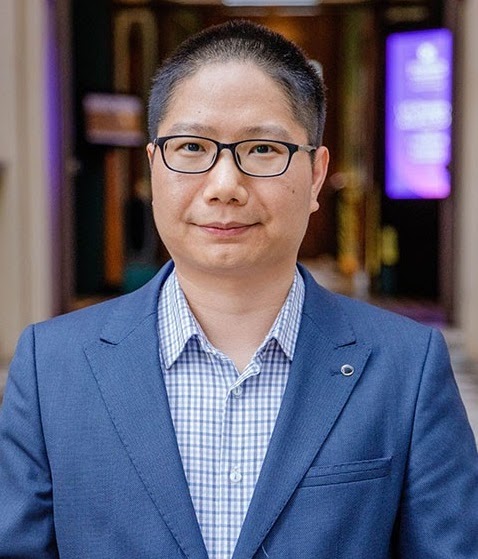}}]{Hongzhi Yin} (Senior Member, IEEE) received a PhD degree in computer science from Peking University, in 2014. He works as an ARC Future Fellow, Full Professor and Director of Responsible Big Data \& Intelligence Lab  at The University of Queensland, Australia. He has made notable contributions to recommendation systems,  graph learning, and decentralized and edge intelligence. He has published 360+ papers with an H-index of 88.
\end{IEEEbiography}

\vspace{0.1pt}
\begin{IEEEbiography}[{\includegraphics[width=1in,height=1.5in,clip,keepaspectratio]{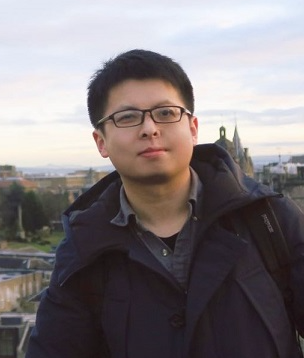}}]{Junliang Yu}
is an ARC DECRA Fellow at Griffith University, specializing in RS, data-centric AI, and graph learning. He has published 30 publications in premier venues, such as KDD, SIGIR, WWW, WSDM, CIKM, TKDE, etc, with five of conference papers being recognized as the most influential papers by Paper Digest and three of journal papers being recognized as ESI hot / highly cited papers.
\end{IEEEbiography}

\vspace{0.1pt}

\begin{IEEEbiography}[{\includegraphics[width=1in,height=1.25in,clip,keepaspectratio]{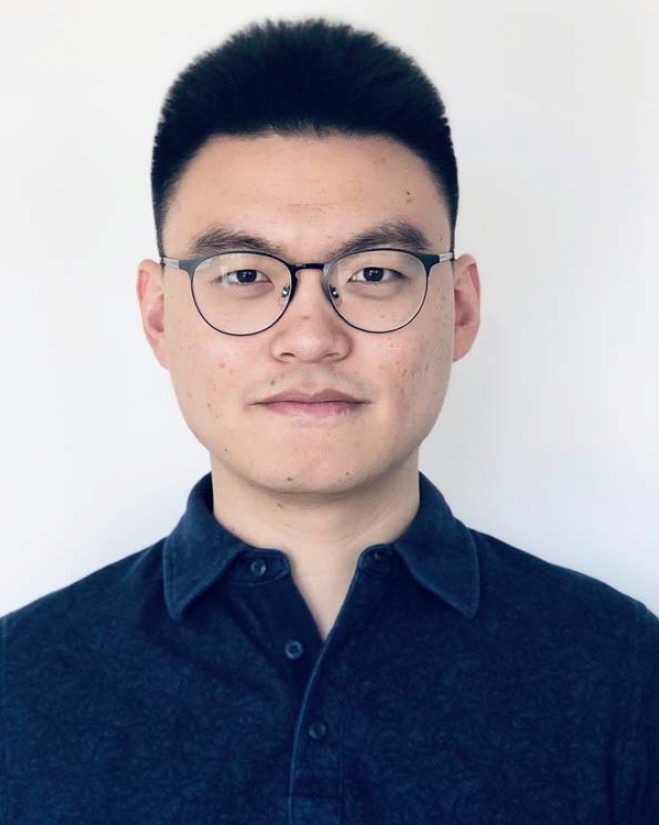}}]{Tong Chen} received the Ph.D. degree in computer
science from The University of Queensland, Brisbane, QLD, Australia, in 2020. He is currently an
Associate Professor with the Data Science Research
Group, School of Electrical Engineering and Computer Science, The University of Queensland. His
research interests include data mining, recommender
systems, user behavior modeling, and predictive analytics.
\end{IEEEbiography}

\vspace{0.1pt}
\begin{IEEEbiography}[{\includegraphics[width=1in,height=1.25in,clip,keepaspectratio]{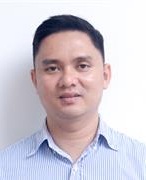}}]{Quoc Viet Hung Nguyen} earned his Master and PhD degrees from EPFL (Switzerland). He works as an Associate Professor at Griffith University. He received Australia Discovery Early Career Researcher Award in 2020. His research focuses on Data Integration, Data Quality, Recommender Systems, and Big Data Visualization, with special emphasis on web data, social data, IoT data and satellite data.
\end{IEEEbiography}

\vspace{0.1pt}

\begin{IEEEbiography}[{\includegraphics[width=1in,height=1.1in,clip,]{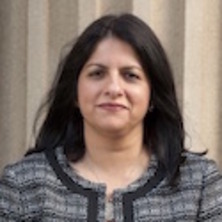}}]{Shazia Sadiq} is a Professor of Computer Science at the School of Information Technology and Electrical Engineering, The University of Queensland. Her research focuses on responsible data management. Her work has contributed to advancing knowledge on data quality management, scalable data curation and cleaning, and bias mitigation for advanced opaque analytical models and techniques.
\end{IEEEbiography}

\vspace{0.1pt}

\begin{IEEEbiography}[{\includegraphics[width=1in,height=1.25in,clip,keepaspectratio]{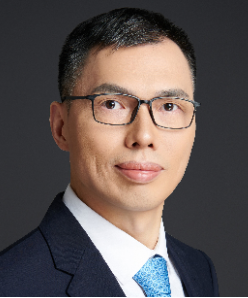}}]{Tianrui Li} is currently a Professor and the
Director of the Key Laboratory of Cloud Computing and Intelligent Techniques, Southwest
Jiaotong University. He serves as Editor-in-Chief
of Human-Centric Intelligent Systems, Editor of
Information Fusion and Associate Editor of ACM
Transactions on Intelligent Systems and Technology. He has authored or coauthored more than 500 research papers in refereed journals (e.g. IEEE TPAMI, IJCV, IEEE TKDE) and conferences (e.g. CVPR, ICCV, KDD). His research interests include big data, cloud computing, data mining, granular computing, and rough sets. He is a Fellow of IRSS and Senior Member of the ACM and IEEE.
\end{IEEEbiography}


 




\vfill

\end{document}